\documentclass[useAMS,usenatbib]{mnras}

\usepackage{url,times,graphicx,hyperref,amsmath,amsfonts,amssymb,aas_macros,color,epsfig,epstopdf}
\usepackage{booktabs}

\def\r1{$r_1$}

\def\m200{$M_{200}$}

\newcommand{\hMpc}{{\ifmmode{h^{-1}{\rm Mpc}}\else{$h^{-1}$Mpc}\fi}}
\newcommand{\hkpc}{{\ifmmode{h^{-1}{\rm kpc}}\else{$h^{-1}$kpc}\fi}}
\newcommand{\hMsun}{{\ifmmode{h^{-1}{\rm {M_{\odot}}}}\else{$h^{-1}{\rm{M_{\odot}}}$}\fi}}
\newcommand{\ltsima}{$\; \buildrel < \over \sim \;$}
\newcommand{\gtsima}{$\; \buildrel > \over \sim \;$}
\newcommand{\lsim}{\lower.5ex\hbox{\ltsima}}
\newcommand{\gsim}{\lower.5ex\hbox{\gtsima}}

\def\lesssim{\mathrel{\hbox{\rlap{\hbox{\lower4pt\hbox{$\sim$}}}\hbox{$<$}}}}
\def\gtrsim{\mathrel{\hbox{\rlap{\hbox{\lower4pt\hbox{$\sim$}}}\hbox{$>$}}}}

\newcommand{\beq}{\begin{equation}}
\newcommand{\eeq}{\end{equation}}
\def\beqa{\begin{eqnarray}}
\def\eeqa{\end{eqnarray}}
\def\hMpc{$h^{-1}\,{\rm Mpc}$}
\def\hkpc{$h^{-1}\,{\rm kpc}$}

\def\head{
 \vbox to 0pt{\vss
                   \hbox to 0pt{\hskip 440pt\rm LA-UR-10-07069\hss}
                  \vskip 25pt}}

\title[Baryonic clues to the diversity of rotation curves]
{Baryonic clues to the puzzling diversity of dwarf galaxy rotation curves}
\author[I. Santos-Santos et al.]
       {Isabel M.E. Santos-Santos$^{1}$\thanks{E-mail:
           isantos@uvic.ca}, Julio F. Navarro$^{1}$,  Andrew
         Robertson$^{2}$,
         \newauthor Alejandro Ben\'itez-Llambay$^{2}$, Kyle A. Oman$^{2}$,
         Mark R. Lovell$^{3,2}$,
         \newauthor Carlos S. Frenk$^{2}$,
 Aaron D. Ludlow$^{4}$ , Azadeh Fattahi$^{2}$, Adam Ritz$^{1}$.
          \\
$^{1}$Department of Physics and Astronomy, University of Victoria, Victoria, BC, Canada V8P 5C2\\
$^{2}$Institute for Computational Cosmology, Department of Physics, Durham University, South Road, Durham, DH1 3LE, UK\\
$^{3}$Center for Astrophysics and Cosmology, Science Institute, University of Iceland, Dunhagi 5, 107 Reykjavik, Iceland\\
$^{4}$International Centre for Radio Astronomy Research, University of Western Australia, 35 Stirling Highway, Crawley, Western Australia 6009, Australia
}

\begin{document}

\date{Accepted XXXX . Received XXXX; in original form XXXX}
\pagerange{\pageref{firstpage}--\pageref{lastpage}} \pubyear{0000}
\maketitle
\label{firstpage}

\begin{abstract}
  We use a compilation of disc galaxy rotation curves to assess the
  role of the luminous component (``baryons'') in the rotation curve
  diversity problem. As in earlier work, we find that rotation curve
  shape correlates with baryonic surface density: high surface density
  galaxies have rapidly-rising rotation curves consistent with cuspy
  cold dark matter halos; slowly-rising rotation curves
  (characteristic of galaxies with inner mass deficits or ``cores'')
  occur only in low surface density galaxies. The correlation,
  however, seems too weak to be the main driver of the diversity. In
  addition, dwarf galaxies exhibit a clear trend, from ``cuspy''
  systems where baryons are unimportant in the inner mass budget to
  ``cored'' galaxies where baryons actually dominate. This trend
  constrains the various scenarios proposed to explain the diversity,
  such as (i) baryonic inflows and outflows during galaxy formation;
  (ii) dark matter self-interactions; (iii) variations in the baryonic
  mass structure coupled to rotation velocities through the ``mass
  discrepancy-acceleration relation'' (MDAR); or (iv) non-circular
  motions in gaseous discs. Together with analytical modeling and
  cosmological hydrodynamical simulations, our analysis shows that
  each of these scenarios has promising features, but none
  seems to fully account for the observed diversity. The MDAR, in
  particular, is inconsistent with the observed trend between rotation
  curve shape and baryonic importance; either the trend is caused by
  systematic errors in the data or the MDAR does not apply. The origin
  of the dwarf galaxy rotation curve diversity and its relation to the
  structure of cold dark matter halos remains an open issue.
\end{abstract} 

\noindent
\begin{keywords}
 galaxies: dwarf -- evolution -- formation -- haloes cosmology: theory -- dark matter
 \end{keywords}


 \section{Introduction}
 \label{SecIntro}

 The non-linear structure of dark matter halos is a 
 solid
 prediction of the Lambda Cold Dark Matter (LCDM) paradigm for
 structure formation. Numerical 
 ``dark-matter-only'' (hereafter, DMO) 
 simulations have consistently shown
 that the density profiles of LCDM halos are approximately
 self-similar, so that the full mass profile of a halo depends on a
 single parameter, such as the virial\footnote{We define the virial
   quantities of a system as those defined by a mean density of
   $200\times$ the critical density for closure. Virial parameters are
   identified by a ``200'' subscript.} mass of the system
 \citep[][hereafter, NFW]{Navarro1996a,Navarro1997}. This prediction
 may be contrasted with observation using the rotation curves of dark
 matter-dominated systems, such as dwarf galaxies.

\begin{figure}
\includegraphics[width=\linewidth]{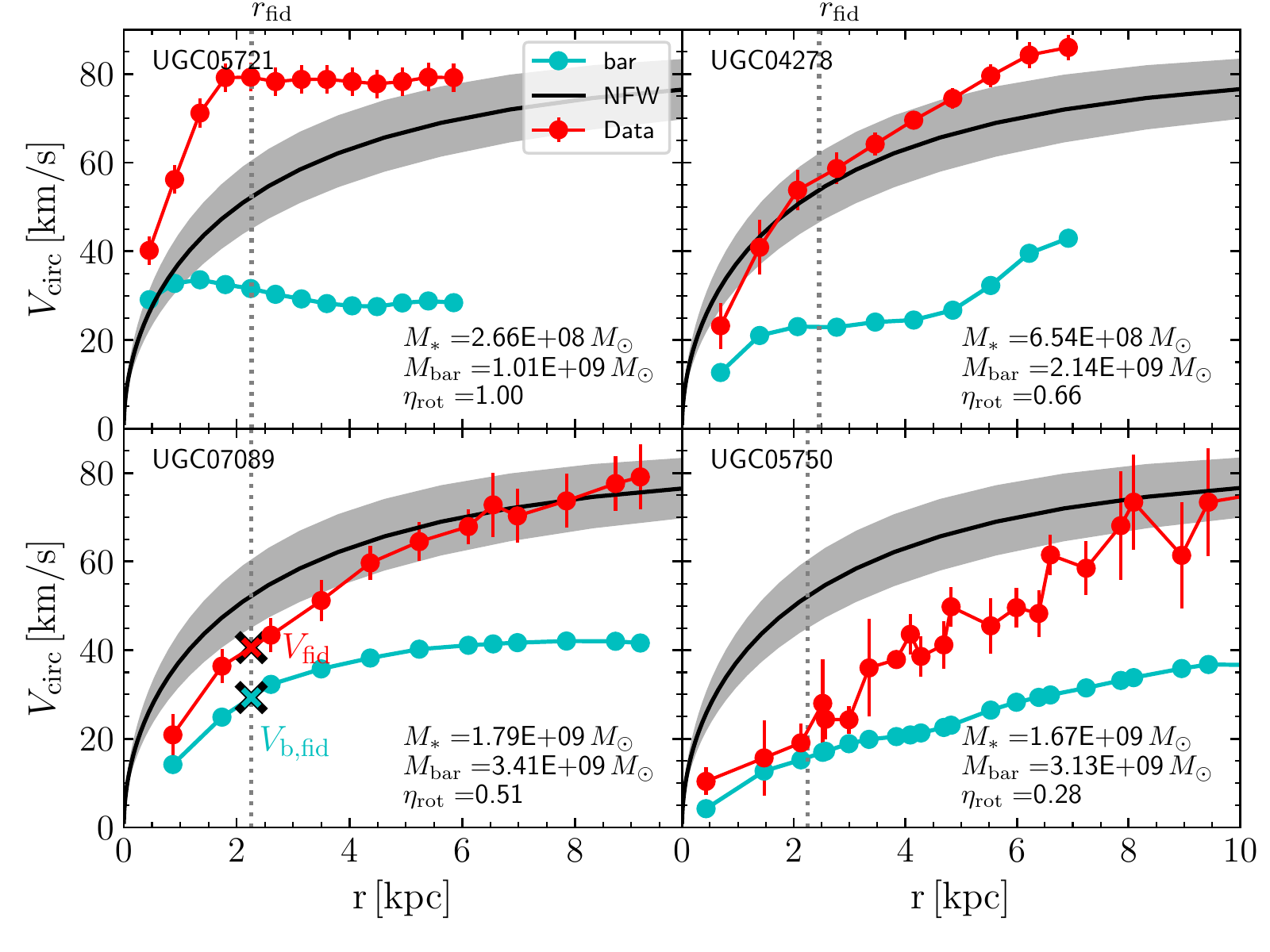}
\centering
\caption{ Examples of rotation curves of dwarf galaxies from the SPARC
  dataset with $V_{\rm max}\sim 80$ km/s. The four galaxies have been
  chosen to span a range of rotation curve shapes, from fast-rising
  (top-left) to slow-rising (bottom-right) relative to the LCDM
  predictions, shown by the black line and gray shaded area. Observed
  rotation speeds are shown in red; the baryonic contribution
  (gas+stars) is shown in cyan. Dotted vertical lines indicate
  $r_{\rm fid}$, the inner fiducial radius adopted in our analysis
  (see eq.~\ref{EqRfid}).  The red and
  cyan crosses in the bottom-left panel illustrate two of the characteristic velocities used in
  our study; the rotation velocity at $r_{\rm fid}$,
  $V_{\rm fid}\equiv V_{\rm rot}(r_{\rm fid})$, and the baryonic
  contribution to the circular velocity ar $r_{\rm fid}$,
  $V_{\rm b,fid}\equiv V_{\rm bar}(r_{\rm fid})$. The total stellar and
  baryonic masses, as well as the rotation curve shape parameter
  $\eta_{\rm rot}=V_{\rm fid}/V_{\rm max}$, are given in the legends
  of each panel. }
\label{FigRCDiv80}
\end{figure}

 In LCDM, dwarf galaxy rotation curve shapes are expected to be nearly
 identical for systems with similar maximum circular velocity,
 $V_{\rm max}$, which is a reliable proxy for the halo virial
 mass. Observed rotation curves, however, deviate from this simple
 prediction, and show great diversity at fixed $V_{\rm max}$. We
 illustrate this in Fig.~\ref{FigRCDiv80}, where the rotation curves
 of $4$ galaxies with $V_{\rm max}\sim 80$ km/s are compared with the
 circular velocity profile of a Navarro-Frenk-White (NFW) halo with
 parameters as expected for a Planck-normalized LCDM cosmology
 \citep{Ludlow2016}.

 All of these galaxies are heavily dark matter dominated in the
 outskirts, where they reach approximately the same $V_{\rm max}$, but
 the shapes of their rotation curves vary greatly. Although UGC
 04278 (top-right panel in Fig.~\ref{FigRCDiv80}) follows roughly the
 expected NFW circular velocity profile, the other three deviate from
 this prediction. Rotation curves that rise more sharply than the NFW
 curve (UGC 05721; top-left) are not unexpected, and may arise, in
 principle, from the accumulation of baryons (i.e., stars plus gas) in
 the inner regions and the ensuing contraction of the halo.

On the other hand, the two galaxies in the bottom panels of
Fig.~\ref{FigRCDiv80} are more problematic, as they have inner
velocities well below the expected values. This implies a sizable
``inner deficit'' of matter relative to LCDM, a feature that is often
associated with a constant-density ``core'' in the dark matter
distribution. These two galaxies are thus clear examples of the
well-known ``cusp-core'' controversy \citep{Moore1994,Flores1994,DeBlok2001,Gentile2004,DeBlok2010},
which, as argued by \citet{Oman2015}, is best characterized as an
inner mass deficit relative to the LCDM predictions. Note that this
deficit affects only {\it some} galaxies, and that others are actually
quite consistent with LCDM, at least according to this measure.

The origin of the diversity illustrated in Fig.~\ref{FigRCDiv80} is
still unclear, and has elicited a number of proposals that are being
actively debated in the literature. These proposals may be grouped
into four broad categories. One is that the diversity is caused by the
effects of baryonic inflows and outflows during the formation of the
galaxy, which lead to gravitational potential fluctuations that may
rearrange the inner dark matter profiles \citep[see;
e.g.,][]{Navarro1996b,Read2005,Mashchenko2006,Brook2012,Governato2012,Pontzen2012,Chan2015}. In
this scenario, ``cores'' are created by feedback-driven blowouts that
remove baryons from the inner regions, leading to a reduction of the
inner dark matter content.  These cores can, in principle, be
reversed, and dark matter cusps may be recreated by subsequent
baryonic (or dark) mass infall  \citep{Laporte2015,Tollet2016,Benitez-Llambay2019}.
This ``baryon-induced cores and cusps'' mechanism (hereafter BICC, for
short) offers in principle an appealing potential explanation for the
observed diversity.

A second scenario argues that dark matter self-interactions are
responsible for ``heating up'' the inner regions of a CDM halo into a
core, thus reducing the central densities and allowing for
slowly-rising rotation curves such as those in the bottom panels of
Fig.~\ref{FigRCDiv80} \citep{Spergel2000}. Galaxies with
rapidly-rising rotation curves are more difficult to accommodate in
this self-interacting dark matter scenario (hereafter SIDM), where
they are ascribed to either systems that were originally so dense that
the resulting core is negligibly small, or to systems where the central
baryonic potential is deep enough to affect the SIDM density profile
\citep[e.g.,][]{Rocha2013,Kaplinghat2016,Kamada2017,Ren2019}.

A third possibility is that the diversity is generated by variations
in the spatial distribution of the baryonic component. Indeed, it has
been argued that galaxy rotation speeds at all radii may be inferred
directly from the baryonic matter distribution via the ``mass
discrepancy-acceleration relation'' \citep[hereafter,
MDAR\footnote{This relation is also known as the ``radial acceleration
  relation", or RAR.},][]{McGaugh2004,McGaugh2016,Lelli2016}. In this scenario the
radial acceleration associated with circular motion,
$g_{\rm obs}(r)=V_{\rm rot}^2(r)/r$, is linked to the baryonic
contribution to such acceleration, $g_{\rm bar}=V_{\rm bar}^2(r)/r$,
through a simple function, $g_{\rm obs}(g_{\rm bar})$, with 
small scatter.  Thus, the diversity in the rotation curve shapes would
result simply from the diverse contribution of baryons to the
acceleration in the inner regions of galaxies with similar $V_{\rm max}$.

Finally, the possibility has been raised that, at least in part, the
diversity may be due to uncertainties in the circular velocities
inferred from observations. The recent work of \citet{Marasco2018} and
\citet{Oman2019} argues that the triaxiality of the dark matter halo
may induce non-circular (i.e., elliptical) closed orbits in 
gaseous discs \citep[see also][]{Hayashi2006}. Depending on how the kinematic principal axes of a particular
galaxy are aligned relative to the major and minor axes of the orbital
ellipses, the inferred velocities may over- or under-estimate the true
circular velocity, sometimes by large amounts. This could also, in
principle, explain the observed diversity. 

How can we tell these scenarios apart? The papers cited in the above discussion have
already shown that each of these mechanisms {\it can}, in principle,
modify the LCDM mass profiles enough to account for the observed
diversity. Therefore, assessing the viability of each of these scenarios
must rely on a more detailed elaboration of their predictions as well
as on the use of ancillary data and diagnostics. This is what we
attempt in the present paper, where we use cosmological hydrodynamical
simulations of galaxy formation in these scenarios, as well as a more
detailed analysis of the role of baryons on different measures of
rotation curve diversity to gauge the success of each of the above
scenarios.

We begin with a brief description of the observational datasets 
(Sec.~\ref{SecObs}), followed by a description of the cosmological
simulations adopted for our analysis (Sec.~\ref{SecSims}). We present
our main results in Sec.~\ref{SecRes}, and summarize our main
conclusions in Sec.~\ref{SecConc}.

As we were preparing this paper for submission, we became aware of a
recent preprint by \citet{Kaplinghat2019}, who analyze many of the
same issues we address here. Some of our conclusions agree with
theirs, others do not. We discuss briefly
similarities and differences in Sec.~\ref{SecConc}.
 
\section{Observational data}
\label{SecObs}

Our compilation of rotation curves from the literature includes
datasets from the Spitzer Photometry \& Accurate Rotation Curves
project \citep[SPARC;][]{Lelli2016}; from The HI Nearby Galaxy Survey
\citep[THINGS;][]{deBlok2008}; from the Local Irregulars That Trace
Luminosity Extremes, The HI Nearby Galaxy Survey \citep[LITTLE THINGS
;][]{Oh2015}; as well as from the work of \citet{Adams2014} and
\citet{Relatores2019}. 
 
All rotation curves in this compilation were inferred from
high-resolution HI and/or H$\alpha$ velocity fields, and include
asymmetric drift corrections when needed. In all cases the velocity
field data has been combined with photometry to construct mass models
that include the stellar, gaseous, and dark matter components. In
particular, the SPARC, THINGS and LITTLE THINGS data make use of
\textit{Spitzer} 3.6$\mu$m surface photometry, while \citet{Adams2014}
and \citet{Relatores2019} use $r$-band images from a variety of
sources. If the same galaxy is common to more than one survey, we
adopt the SPARC data, because the majority of galaxies in our sample
come from that compilation.
 
To minimize the inclusion of rotation curves that might be affected by
substantial uncertainty we only consider galaxies with inclinations
$i>30^{\circ}$, and omit the ``grade=3'' galaxies of
\citet{Relatores2019}. (We refer the reader to that paper for
details.)  Furthermore, as our analysis relies on comparing rotation
velocities in the inner and outer regions, we retain only systems
whose rotation curves cover a relatively wide radial range. More
specifically, we retain only systems where the last measured point of
the rotation curve ($r_{\rm last}, V_{\rm last}$) is at least twice as
far from the centre as a ``fiducial'' inner radius, defined as
\begin{equation}
   r_{\rm fid} = 2 (V_{\rm max}/70\, {\rm km/s}) \, {\rm kpc}.
\label{EqRfid}
\end{equation}

We note that in most cases $V_{\rm last}\approx V_{\rm max}$, and that
the scaling of $r_{\rm fid}$ with $V_{\rm max}$ ensures that the ratio
$\eta_{\rm rot}=V_{\rm fid}/V_{\rm max}$ (where
$V_{\rm fid}\equiv V_{\rm rot}(r_{\rm fid})$) is a simple but reliable
indicator of the shape of the rotation curve for dwarf and massive
galaxies alike. Rapidly-rising rotation curves have high values of
$\eta_{\rm rot}$, approaching unity for rotation curves that remain
approximately flat from the inner to the outermost regions.

Because sharply-rising rotation curves are expected from cuspy dark
matter profiles, we shall at times loosely refer to rotation curves with
$\eta_{\rm rot}\sim 1$ as ``cuspy''. On the other hand, systems
with $\eta_{\rm rot} \ll 1$ have very slowly-rising rotation curves,
consistent with ``cores''. We shall occasionally refer to such systems
as having ``cored'' mass profiles or ``cored'' rotation curves.

Our compilation retains a total of 160 galaxies, spanning a wide
range in $V_{\rm max}$ (from $\sim 20$ to $\sim 380$ km/s ) and in
stellar mass (from $M_{\rm star}\sim 1.6\times 10^6\, M_\odot$ to
$\sim 2.5 \times 10^{11} \, M_\odot$). Our analysis makes use of
published mass models, which include the combined gravitational effect
of gas and stars --which we shall hereafter refer to as ``baryons''--, on
the rotation curve. In practice, we shall use
$V_{\rm bar}^2(r)=V_{\rm gas}^2(r)+V_{\rm stars}^2(r)$, where the
latter two terms are the contributions to the circular velocity of gas
and stars reported in the literature. We have also computed baryonic
half-mass radii, $r_{\rm b,half}$, assuming spherical symmetry (i.e.,
$M_{\rm bar}(<r)=r V_{\rm bar}^2(r) /G$) and that the baryonic
component does not extend beyond $r_{\rm last}$.

When necessary, we estimate virial masses, $M_{200}$, for each system
assuming an NFW profile of the same maximum circular velocity and a
concentration parameter, $c$, taken from \citet{Ludlow2016}'s median
$M_{200}(c)$ relation.  We list, for each galaxy in our sample, the
specific structural and velocity parameters used in our analysis in
Table \ref{TabObsData}. Although this compilation contains most
galaxies with high-quality rotation curves inferred from 2D velocity
fields, it is important to note that our sample may be subject to substantial
selection biases that are not easy to quantify. We shall hereafter
assume that these galaxies are representative of the galaxy population
as a whole, but this is an assumption that may require revision once
better, more complete datasets become available.

\section{Numerical simulations}
\label{SecSims}

\subsection{LCDM simulations: EAGLE/APOSTLE}
\label{SecAPOSTLE}

The APOSTLE project is a set of twelve zoom-in simulations of ``Local
Group''-like regions selected from a 100$^3$ Mpc$^3$
DMO cosmological box run in a WMAP-7
cosmology.  These Local Group (LG) regions are defined by the presence
of a pair of halos that meet mass, relative velocity, and isolation
criteria that match observed constraints on the Milky Way-Andromeda
pair \citep{Fattahi2016,Sawala2016}. These LG volumes have been run at
three different levels of resolution. We shall use for this analysis
the highest-resolution set (labelled ``AP-L1''), with particle masses
$m_{\rm dm}\sim 5\times10^4\,M_\odot$, $m_{\rm gas}\sim 10^4\,M_\odot$
and a maximum physical gravitational softening length of $134$ pc. Our
analysis will use isolated (i.e., not satellites) systems found at
$z=0$ within $\sim 2.5$ Mpc from the barycentre of the two main
galaxies in each volume.

The APOSTLE simulations were run with the EAGLE (Evolution and
Assembly of GaLaxies and their Environments) galaxy formation code
\citep{Schaye2015, Crain2015}, which includes radiative cooling, star
formation, stellar feedback, black hole growth and active galactic
nuclei (AGN) feedback (the latter negligible for LG galaxies).  In
particular, star formation assumes a metallicity-dependent density
threshold \citep{Schaye2004} 
of the form
\begin{equation}
n_{\rm thr} =\rm min\left[ n_{\rm thr,0} \left( {Z/Z_0}\right)^{-\alpha},n_{\rm max} \right],
\label{EqSFThr}
\end{equation}
where $n_{\rm thr,0}=0.1$ cm$^{-3}$, $n_{\rm max}=10$ cm$^{-3}$,
$Z_0=0.002$, and $\alpha=0.64$.
Stellar feedback mimicking the
effects of stellar winds, radiation pressure and supernova explosions
is accounted for using a stochastic, thermal prescription
\citep{DallaVecchia2012}.
 
\subsection{LCDM simulations: NIHAO}
\label{SecNIHAO}

The NIHAO (Numerical Investigation of a Hundred Astrophysical Objects)
project is a set of $\sim$100 cosmological zoom-in hydrodynamical
simulations of isolated galaxies performed using the ESF-Gasoline2
code \citep{Wadsley2004, Wang2015} and run in a flat LCDM cosmology
with parameters from \citet{Planck2014}.  These simulations span a
wide range of halo virial masses, from $\sim 5 \times 10^9$ to
$2\times 10^{12} \,M_\odot$. All NIHAO have spatial resolution high
enough to resolve the mass profile of all systems reliably down to
$1\%$ of the virial radius. Particle masses scale with halo virial
mass so that all halos are resolved with similar numbers of
particles. As an example, they are
$m_{\rm dm}\sim 2\times10^4\,M_\odot$ and
$m_{\rm gas}\sim 3.5\times10^3\,M_\odot$ for a 10$^{10}\, M_\odot$
halo.

Subgrid physics include a recipe for star formation that matches the
Kennicutt-Schmidt Law in regions with a temperature below $15000$ K
and density above a threshold, $n_{\rm thr}> 10.3$ cm$^{-3}$.  The
algorithm includes stellar feedback from supernovae \citep[implemented
through a blast-wave formalism;][]{Stinson2006} and from massive stars
prior to their explosion as SNe \citep[``early stellar
feedback";][]{Stinson2013}.

The star formation and feedback algorithms are such that NIHAO dark
matter halos can expand to form cores, with the degree of expansion
depending mainly on the stellar-to-halo mass ratio
\citep{Tollet2016,Dutton2016,Dutton2019b}.  As a result, NIHAO
galaxies show a fairly wide diversity of rotation curves, as discussed
by \citet{Santos-Santos2018}.

\subsection{LCDM simulations: EAGLE-CHT10}
\label{SecCHT10}

This simulation series, first presented in
\citet{Benitez-Llambay2019}, evolves cosmological boxes $12$ Mpc on a
side, and are run with the same code and cosmology as the
EAGLE/APOSTLE project (see Sec.~\ref{SecAPOSTLE}). The main difference
is that they adopt a higher threshold for star formation, independent
of metallicity. In this paper we consider a run with constant
$n_{\rm thr}=10$ cm$^{-3}$, which is roughly $100\times$ higher than
that used in APOSTLE.  Mass resolution is given by
$m_{\rm dm}\sim 4\times10^5\,M_\odot$ and
$m_{\rm gas}\sim8\times10^4\,M_\odot$. As reported in
\citet{Benitez-Llambay2019}, a higher star formation threshold allows
gas to collapse and become gravitationally dominant at the centre of a
halo. Baryonic outflows driven by supernova feedback are then able to
modify the inner DM density profile, just as in the NIHAO simulations,
which adopt a similar value of $n_{\rm thr}$.

\subsection{SIDM simulations: SIDM10}
\label{SecSIDMSim}

We use two re-simulations of one of the APOSTLE volumes 
\citep[AP01-L1 in the notation of][]{Fattahi2016}.
 One re-simulation is
dark-matter-only, the other includes the same subgrid physical
treatment of star formation and feedback as APOSTLE, except that it
uses the EAGLE-Recal model parameters rather than the EAGLE-Ref
parameters that were used in the APOSTLE runs \citep{Lovell2020}. 
Mass and spatial resolution are the same as in that series
(Sec.~\ref{SecAPOSTLE}). The only difference is that the EAGLE code
has been modified to include a collisional term for dark matter
particle pairwise interactions in order to model the effects of a
$s_{\rm si}=\sigma_{\rm SIDM}/m$=10 cm$^2$/g velocity-independent
self-interaction cross section. The code modifications are described
in detail in \citet{Robertson2017,Robertson2018}.

\begin{figure}
\includegraphics[width=\linewidth]{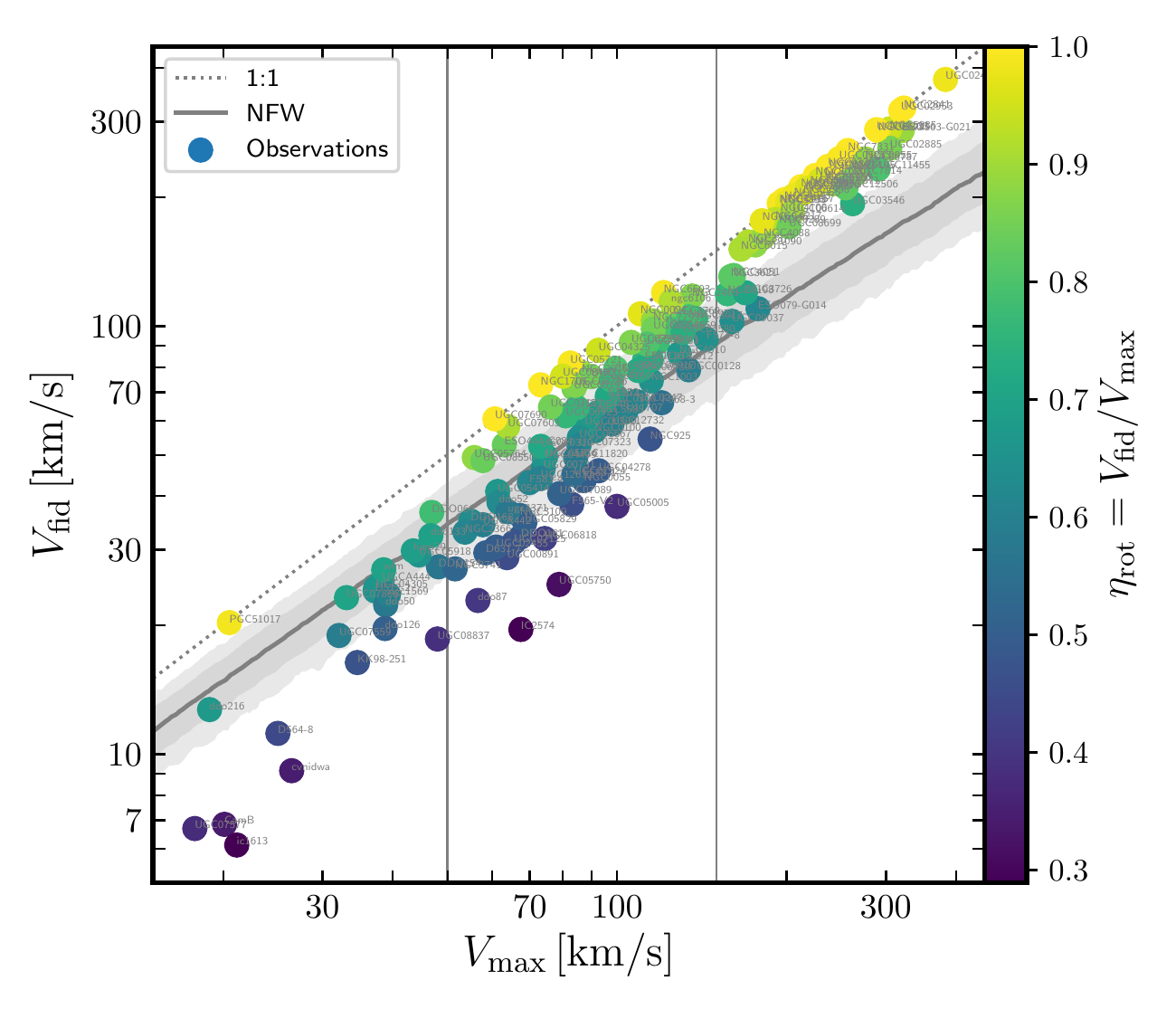}
\centering
\caption{Rotation speed at the inner fiducial radius vs
  maximum rotation speed for all galaxies in our observational
  sample. Galaxies with rapidly-rising rotation curves lie near the
  1:1 dotted line. Rotation curves with ($V_{\rm fid}$,$V_{\rm max}$)
  values consistent with dark matter-only LCDM halos lie along the
  gray curve and shaded area, computed using the median $M_{200}(c)$
  relation (plus 10-90 and 1-99 percentiles) for a Planck-normalized
  cosmology \citep{Ludlow2016}. Slowly-rising rotation curves (i.e.,
  ``cored'' galaxies with a substantial inner mass deficit relative to
  LCDM) lie below the grey-shaded area. Each galaxy is labelled with
  its name and coloured according to $\eta_{\rm rot}$, which we adopt
  as a simple measure of rotation curve shape.}
\label{FigVfidVmax}
\end{figure}

\begin{figure}
\includegraphics[width=\linewidth]{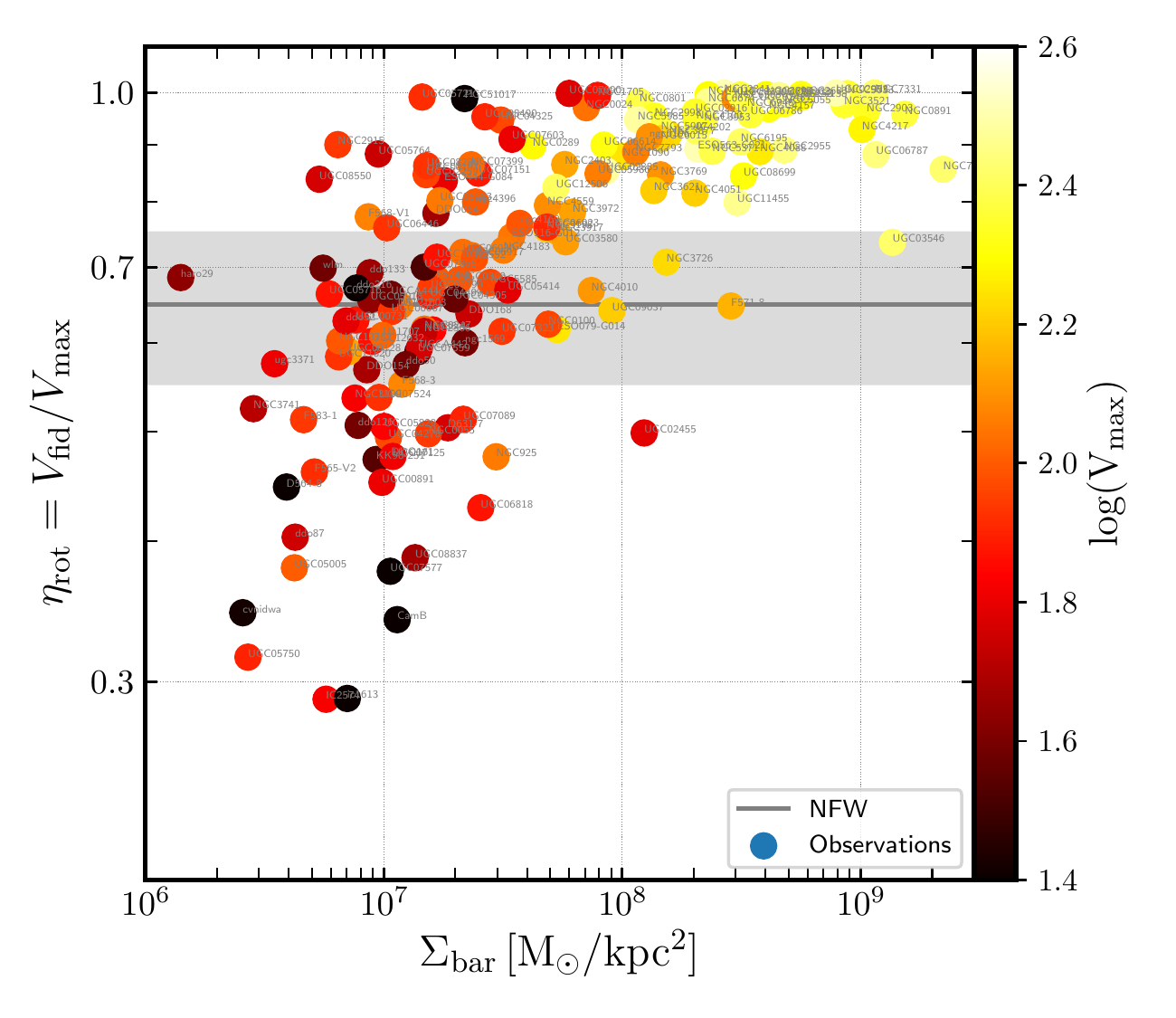}
\centering
\caption{ Rotation curve shape parameter, $\eta_{\rm rot}$, vs
  effective baryonic surface density, $\Sigma_{\rm bar}$. The
  correlation between the two reflects the $\eta_{\rm rot}$ dependence
  on galaxy mass (or $V_{\rm max}$), which in turn correlates strongly
  with surface density. Massive galaxies (high $\Sigma_{\rm bar}$) do
  not show evidence for cores, which only occur in dwarfs, which have
  low $\Sigma_{\rm bar}$. In the dwarf galaxy regime the
  $\eta_{\rm rot}$-$\Sigma_{\rm bar}$ is actually quite weak and
  unlikely to be the cause of the diversity. See text for further
  discussion. }
\label{FigEtaRotSB}
\end{figure}

\section{Results}
\label{SecRes}

\subsection{Rotation curve diversity: Observational results}
\label{SecRCdiv}

\subsubsection{Rotation curve shape vs mass}

The parameter $\eta_{\rm rot}=V_{\rm fid}/V_{\rm max}$ is a useful
measure of the shape of a rotation curve. It contrasts the measured
rotation speeds in the inner regions of the galaxy ($V_{\rm fid}$)
with the maximum rotation speed ($V_{\rm max}$), which is generally
similar to the velocity at the outermost measured point,
$V_{\rm last}$. We note that in some cases the rotation curve may
still be rising at the last measured point, in which case
$V_{\rm max}$ may be underestimated. This, however, should have a
relatively minor effect on $\eta_{\rm rot}$ because a change in
$V_{\rm max}$ would lead to a change in the inner fiducial
radius, $r_{\rm fid}$ (see eq.~\ref{EqRfid}). Recall as well that we only retain systems
where $r_{\rm last}>2\, r_{\rm fid}$, which should minimize any bias
introduced by this effect.

The scaling of $r_{\rm fid}$ with $V_{\rm max}$ means that our measure
of the inner rotation curve adjusts to the total mass of a
system. This is preferable to using a fixed physical radius, and
allows for a proper comparison of the inner and outer regions of the
rotation curve of a galaxy, regardless of mass. For LCDM halos,
$r_{\rm fid}$ is in the rising part of the circular velocity curve,
and one would expect a roughly constant value of
$V_{\rm fid}/V_{\rm max}\sim 0.65$.  We plot these two parameters for
all galaxies in our sample in Fig.~\ref{FigVfidVmax}.

Fig.~\ref{FigVfidVmax} illustrates a few interesting points. One is
that the evidence for ``cores'', defined by an inner mass deficit
relative to LCDM at $r_{\rm fid}$, affects only a fraction of all
galaxies and is restricted to dwarf systems. Indeed, no galaxy with
$V_{\rm max}>150$ km/s is found below the grey shaded band that tracks
the expected loci of dark matter halos in LCDM. Most massive galaxies
closely hug the 1:1 line, suggesting rotation curves that rise
actually more rapidly than expected for LCDM halos and stay flat out
to their last measured radius.  As we shall discuss below
(Sec.~\ref{SecRCBarDom}), this is due to the effect of baryons, which
accumulate at the centre and drive the circular velocity at
$r_{\rm fid}$ to higher values than expected from the dark matter
alone.

Galaxies with $V_{\rm max}<150$ km/s, on the other hand, show a wide
diversity of rotation curve shapes, from rapidly-rising, nearly flat
rotation curve galaxies near the dotted 1:1 line, to very
slowly-rising curves well below the grey band. The latter are systems
where the evidence for an inner mass deficit, or a ``core'', is 
most compelling.

As discussed in Sec.~\ref{SecIntro}, four different scenarios have
been proposed to explain the rotation curve diversity illustrated in
Fig.~\ref{FigVfidVmax}; identifying which one is most consistent with
existing data is the main goal of this paper. Since baryons play an
important role in several of the proposed scenarios, it is important to
check how the diversity correlates with the gravitational contribution
of baryons. We discuss this next.

\subsubsection{Rotation curve shape and baryon surface density}

The role of baryons in determining the shape of the rotation curve has
long been predicated on the basis that the shape of the inner rotation
curve seems to correlate with galaxy surface brightness
\citep[e.g.,][]{DeBlok1996,Swaters2009,Swaters2012,Lelli2013}. We shall use here baryonic
surface density rather than (stellar) surface brightness since baryons
are mainly in gaseous form in many galaxies of our sample.  We explore
this in Fig.~\ref{FigEtaRotSB}, where we plot the rotation curve shape
parameter, $\eta_{\rm rot}$, as a function of the ``effective''
baryonic surface mass density,
$\Sigma_{\rm bar}=M_{\rm bar}/2\pi r_{\rm b,half}^2$. ($M_{\rm bar}$
and $r_{\rm b,half}$ are the total baryonic mass and half-mass radius
of a galaxy, respectively.)

There is indeed a correlation between $\eta_{\rm rot}$ and
$\Sigma_{\rm bar}$, but it is largely a reflection of the galaxy
mass-surface density relation: massive galaxies (none of which have
cores, as discussed in the previous subsection) have higher surface
density than dwarfs (some of which have cores and others which do
not). Aside from this overall trend, when considering only dwarfs
(i.e., $V_{\rm max}<150$ km/s) Fig.~\ref{FigEtaRotSB} shows that the
correlation between $\eta_{\rm rot}$ and $\Sigma_{\rm bar}$ is rather
weak. Although all ``cored'' galaxies (i.e.,
$\eta_{\rm rot}\lsim 0.55$) are low
surface density systems, the converse is not true: there are indeed a number of low
surface density galaxies with ``cuspy'' rotation
curves. Baryonic surface density alone is thus not a reliable indicator of the
presence of a core or cusp in a dwarf galaxy and, therefore, unlikely
to be the origin of the diversity. This is an issue to which we will
return repeatedly throughout our discussion below.

\begin{figure}
\includegraphics[width=\linewidth]{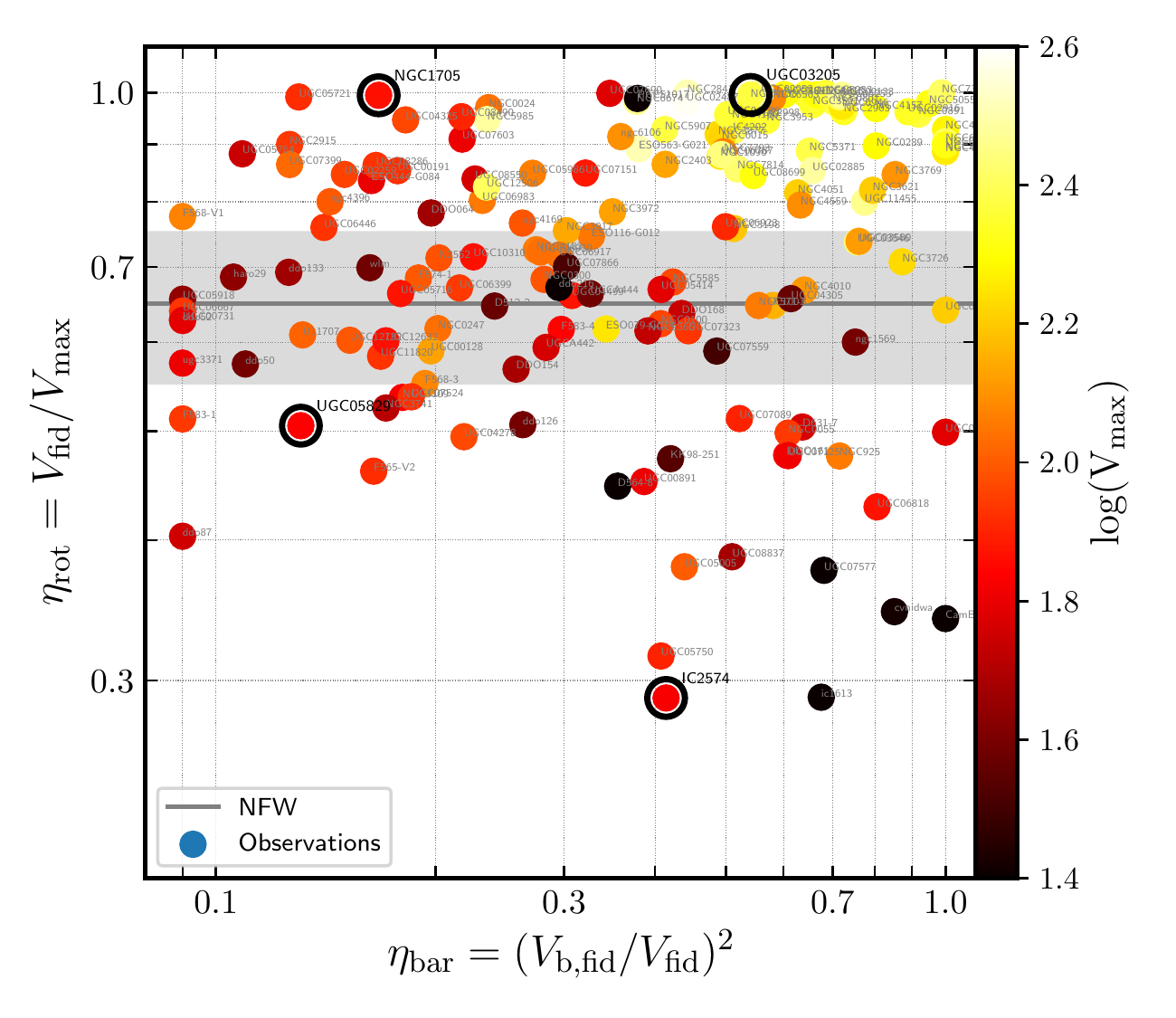}
\centering
\caption{Rotation curve shape parameter, $\eta_{\rm rot}$, as a
  function of the gravitational importance of the baryonic component
  at the inner fiducial radius, $\eta_{\rm bar}$. The latter is 
  approximately the ratio between baryonic and total enclosed masses
  within $r_{\rm fid}$. Galaxies are coloured by maximum circular
  velocity, as indicated by the colourbar. Dark-matter-only LCDM halos
  have, on average, $\eta_{\rm rot}\sim 0.65$, with $10$:$90$ percentile scatter as indicated
  by the grey band. The rotation curves of the four galaxies
  highlighted with black circles are shown in Fig.~\ref{FigRCDiv}. Systems with
  $\eta_{\rm bar}<0.09$ are shown at that value for clarity.}
\label{FigEtaRotEtaBar}
\end{figure}

\subsubsection{Rotation curve shape and baryon central dominance}
\label{SecRCBarDom}

Although, overall, dwarf galaxy rotation curve shapes correlate only
weakly with the effective baryon surface density, it is possible that
baryons shape rotation curves through their role in setting the inner
gravitational potential.  We explore this in
Fig.~\ref{FigEtaRotEtaBar}, where we plot $\eta_{\rm rot}$ vs
$\eta_{\rm bar}$, where the latter is defined as
$\eta_{\rm bar} \equiv (V_{\rm bar}(r_{\rm fid})/V_{\rm fid})^2$.  (We
hereafter define $V_{\rm b,fid}\equiv V_{\rm bar}(r_{\rm fid})$ for
ease of notation.) The (squared) ratio between the baryonic
contribution and the measured rotation velocity at $r_{\rm fid}$ is
roughly equivalent to the ratio between the enclosed baryonic and
total mass within the inner fiducial radius, $r_{\rm fid}$.

\begin{figure}
\includegraphics[width=\linewidth]{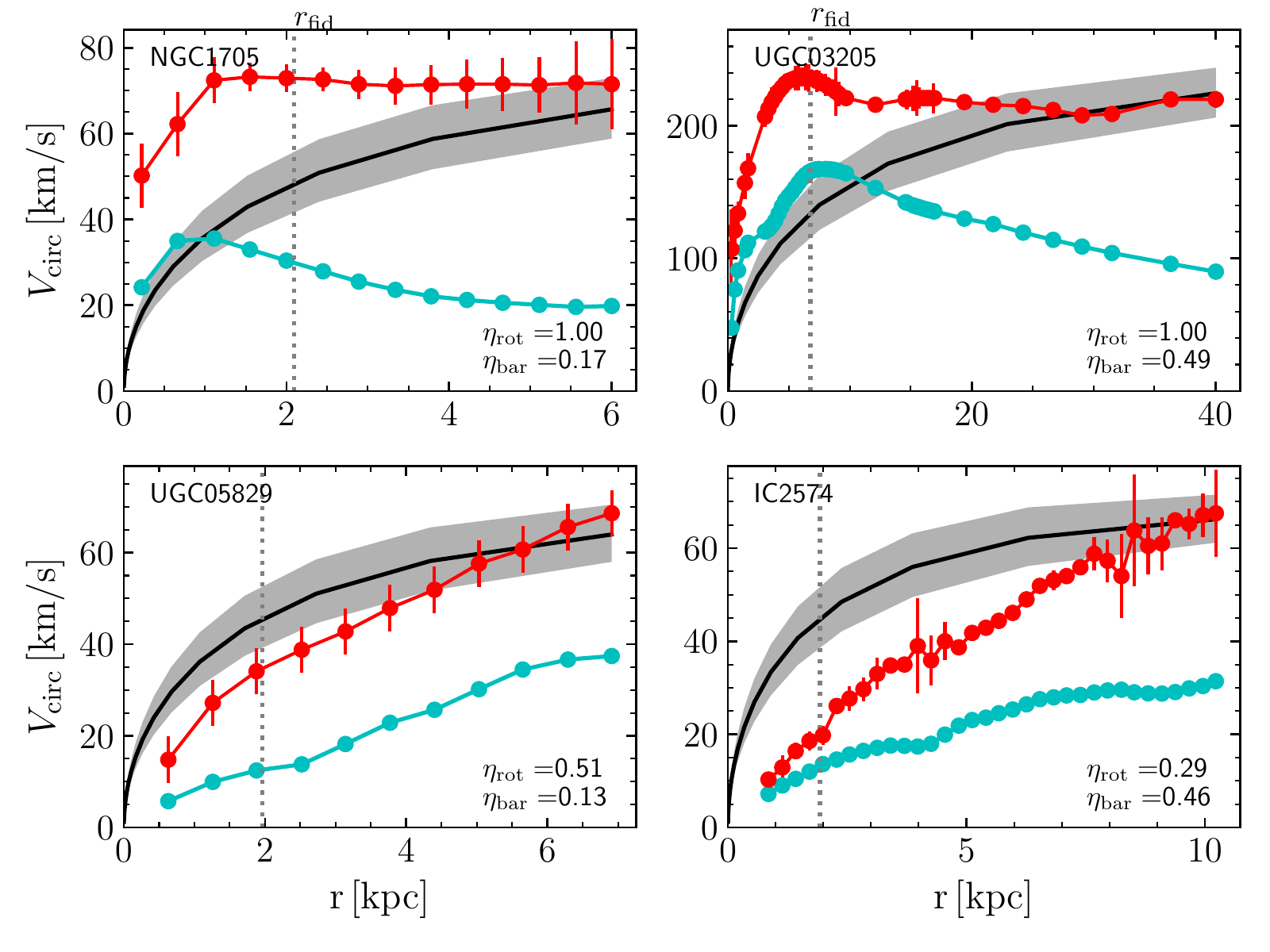}
\centering
\caption{Rotation curves of four galaxies in different regions of
  Fig.~\ref{FigEtaRotEtaBar}, where they are marked with black
  circles. Symbols, colours, and lines are as in
  Fig.~\ref{FigRCDiv80}. The inner fiducial radius, $r_{\rm fid}$, is
  indicated by the vertical dashed line in each panel. Top panels show
  galaxies with rapidly-rising rotation curves, where baryons play an
  important (right) or negligible (left) role at $r_{\rm fid}$. Bottom
  panels are ``cored'' galaxies with slowly-rising rotation curves,
  where, as in top, baryons are gravitationally important (right) or
  negligible (left) at $r_{\rm fid}$. }
\label{FigRCDiv}
\end{figure}

Fig.~\ref{FigEtaRotEtaBar} illustrates a few interesting points. The first
is that in massive disks (i.e., $V_{\rm max}>150$ km/s) rotation
curves are approximately flat (i.e., $\eta_{\rm rot}\sim 1$) and baryons, as
expected, play an important role (i.e., $\eta_{\rm bar}\gsim 0.4$-$0.5$). An
example of this kind of galaxy (UGC 03205) is shown in the top-right panel of Fig.~\ref{FigRCDiv}. The
rotation curve in this case rises actually more rapidly than expected for an LCDM halo of
the same $V_{\rm max}$ (black curve in the same panel) because of the gravitational
importance of the baryons at $r_{\rm fid}$.

For less massive systems the interpretation is less clear: most dwarf
galaxies (defined as having $V_{\rm max}<150$ km/s) scatter from the
top-left to bottom-right corners in Fig.~\ref{FigEtaRotEtaBar}, a
surprising trend for scenarios that envision the importance of baryons
as the main driver of the rotation curve diversity.

Indeed, take, for example, systems at the top-left corner of
Fig.~\ref{FigEtaRotEtaBar}: these are galaxies with rapidly-rising
(``cuspy'') rotation curves but where baryons play a negligible role
at the inner fiducial radius $r_{\rm fid}$.  An example (NGC 1705) is shown in
the top-left panel of Fig.~\ref{FigRCDiv}. As we shall see below, systems like this are
difficult to reproduce in scenarios like BICC and SIDM, where all or
most halos have cores and cuspy rotation curves are assumed to occur
only in systems where baryons dominate the central potential.

A similar comment applies to galaxies at the bottom-right corner of
Fig.~\ref{FigEtaRotEtaBar}: these galaxies have the largest ``cores'',
which, in the BICC scenario, would correspond to systems that have
suffered the effects of explosive baryonic outflows, and where few
baryons should have remained in the galaxy. An example (IC 2574) is
shown in the bottom-right panel of Fig.~\ref{FigRCDiv}. In this
galaxy, as well as in most systems with the largest cores, baryons are
actually as dominant at $r_{\rm fid}$ as in massive, high surface
brightness disks (i.e., $\eta_{\rm bar}>0.5$).

This discussion illustrates how Fig.~\ref{FigEtaRotEtaBar} provides a
useful tool to judge the viability of the various scenarios that aim
to explain the rotation curve diversity. In other words, it is not
enough to identify a mechanism that may modify the inner regions of a
halo to create diversity in the rotation curves of dwarf galaxies; the
same mechanism must also allow galaxies to exhibit the observed
diversity in the importance of baryons in the inner regions and must
reproduce the trends in the $\eta_{\rm rot}$-$\eta_{\rm bar}$ plane
shown in Fig.~\ref{FigEtaRotEtaBar}. This is the key argument of the
analysis that follows, where we shall examine, in turn, the successes
and shortcomings of each of the four scenarios identified in
Sec.~\ref{SecIntro}.
 
\begin{figure*}
\includegraphics[width=0.7\linewidth]{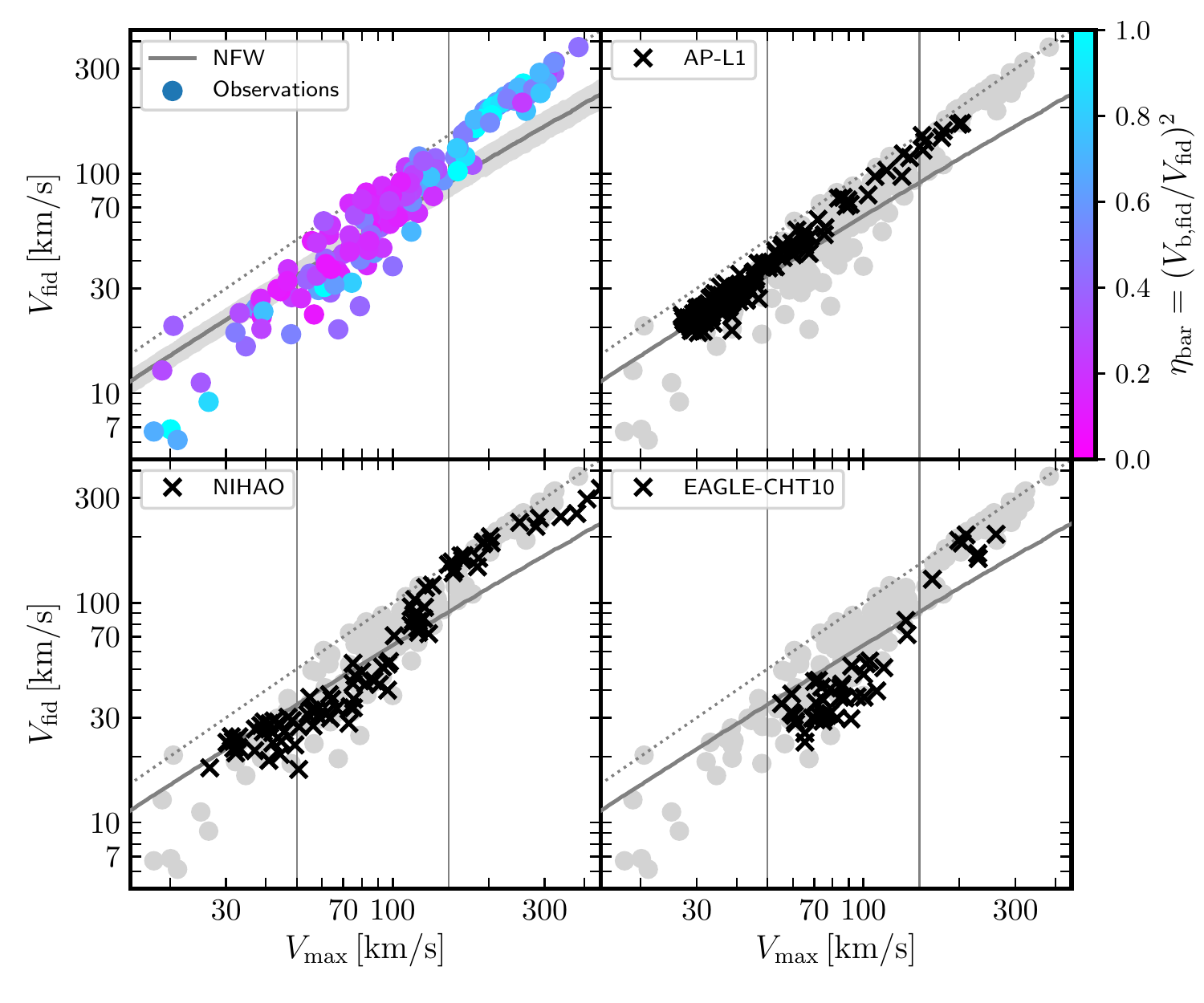}
\centering
\caption{Rotation velocity at the inner fiducial radius vs maximum
  circular velocity, for observations and LCDM simulations. Symbols
  and lines in each panel are as Fig.~\ref{FigVfidVmax}. Top-left
  panel presents the observational sample, coloured by
  $\eta_{\rm bar}$, which measures the gravitational importance of
  baryons at $r_{\rm fid}$. Other panels indicate the results of
  simulations (crosses), but also include the observations, for reference,
  as grey circles. Legends in each panel identify the simulation
  series. Simulated galaxies are only included if the
  \citet{Power2003} convergence radius is smaller than $r_{\rm
    fid}$. See discussion in text.}
\label{FigVfidVmaxLCDM}
\end{figure*}

\begin{figure*}
\includegraphics[width=0.7\linewidth]{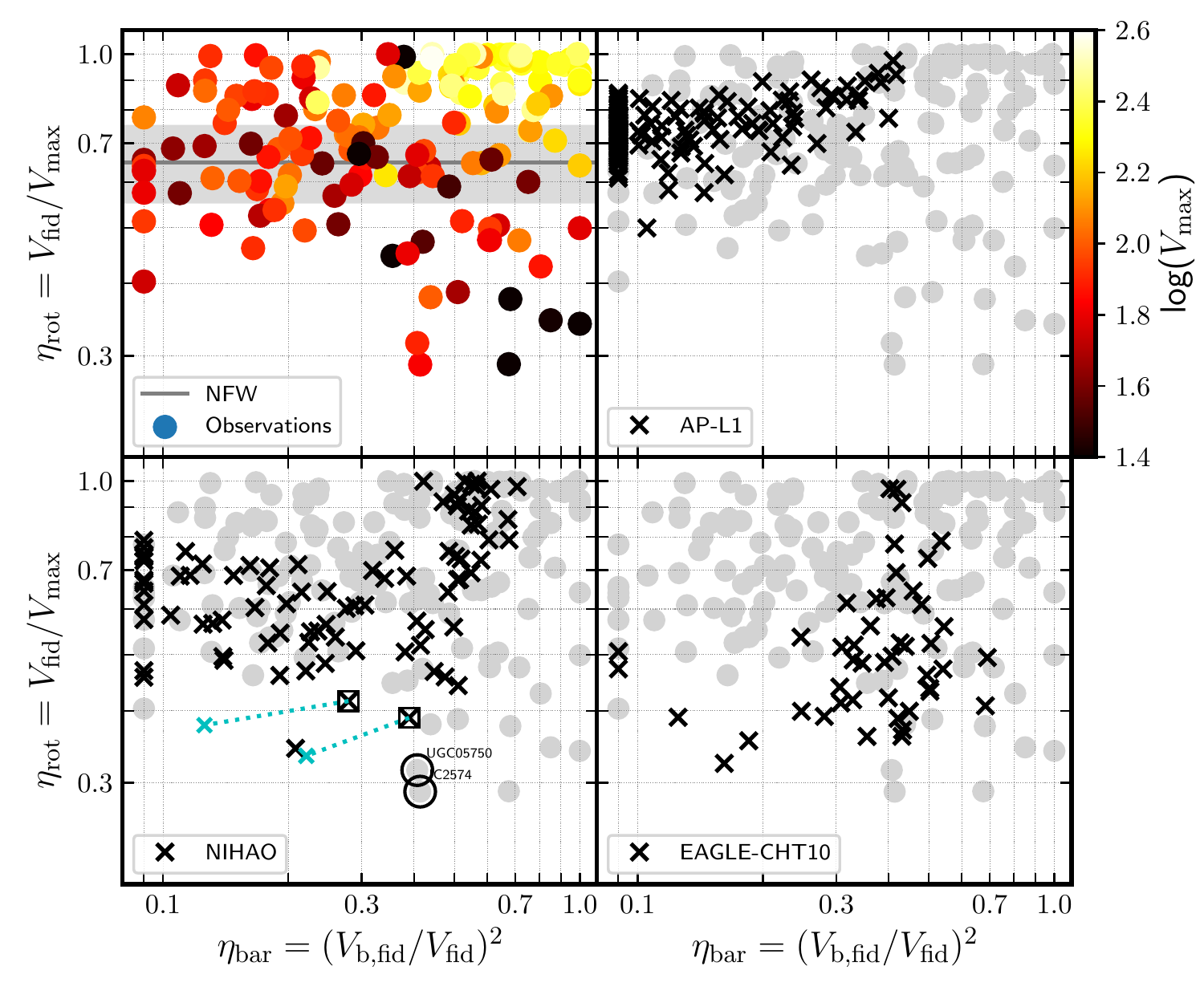}
\centering
\caption{Rotation curve shape parameter, $\eta_{\rm rot}$, vs baryonic
  importance parameter, $\eta_{\rm bar}$, for observations and LCDM
  simulations. Symbols and lines in each panel are as in
  Fig.~\ref{FigEtaRotEtaBar}. Top-left panel presents the
  observational sample, coloured by the maximum circular velocity,
  $V_{\rm max}$. Other panels indicate the results of simulations
  (crosses), but also include observations, for reference, as grey
  circles. Legends in each panel identify the simulation
  series. 
  Systems with $\eta_{\rm bar}<0.09$ are shown
  at that value for clarity. See text in Sec.~\ref{SecNIHAODisc} for a
  discussion of the systems highlighted in the bottom-left panel.  }
\label{FigEtaRotEtaBarLCDM}
\end{figure*}

\subsection{Diversity and BICC (baryon-induced cores/cusps)}
\label{SecDivLCDM}

We shall use several LCDM cosmological hydrodynamical simulations to
compare with observed rotation curves. The first corresponds to
simulations from the APOSTLE project \citep{Sawala2016}, which used
the code developed for the EAGLE project \citep{Schaye2015,Crain2015}
to simulate volumes selected to resemble the Local Group.

Rotation curves from these simulations have been analyzed in a number
of papers \citep[e.g.,][]{Oman2015,Sales2017,Bose2019}, who report that the
inner cuspy structure of the halos is largely unaltered by the
assembly of the galaxy, except for some halo contraction caused by the
accumulation of baryons at the centre \citep[see
also][]{Schaller2015}. No dark matter ``cores'' are formed in APOSTLE,
in the sense that there is no obvious reduction in the inner dark
matter content compared to what would be expected from a DMO
simulation. We shall use galaxies selected from the
``high-resolution'' APOSTLE volumes, referred to as AP-L1 for short
\citep{Fattahi2016}.

The lack of cores in the EAGLE/APOSTLE simulations has been traced to
the relatively low (minimum) gas density threshold for star formation
adopted in that code (Eq.~\ref{EqSFThr}). This prevents the gas that
condenses at the centres of dark matter halos from dominating
gravitationally, minimizing the effects that baryonic inflows and
outflows may have on the dark matter \citep{Pontzen2012}.  Cores do
form in simulations run with the same EAGLE code but with a raised
threshold \citep{Benitez-Llambay2019}. For that reason, we
shall also use here results
from a simulation with $n_{\rm thr}=10$ cm$^{-3}$, labelled
EAGLE-CHT10. 

Dark matter cores have also been reported in simulations from the
NIHAO project \citep{Wang2015}, a series of zoom-in resimulations of
galaxies spanning a wide range in mass. Like EAGLE-CHT10, these
simulations adopt a high star formation threshold, but a rather different
implementation of the star formation and feedback algorithms. 
Details for this and other simulations may be found in
Sec.~\ref{SecSims} and references listed there.

How well do each of these simulation series reproduce the observed
diversity and, importantly, the relation between rotation curve shape
and baryonic importance? We contrast the $V_{\rm fid}$-$V_{\rm max}$
results with observational data in
Fig.~\ref{FigVfidVmaxLCDM}. Simulated galaxies are only included if
the \citet{Power2003} convergence radius is smaller than
$r_{\rm fid}$. Except when otherwise explicitly noted, we shall
estimate circular velocities in simulated galaxies assuming spherical
symmetry; i.e., $V_{\rm circ}^2(R)=GM(<r)/r$.

The top-left panel reproduces the observational data
presented in Fig.~\ref{FigVfidVmax}, coloured this time by the baryon
importance parameter, $\eta_{\rm bar}$. The same observational data is
reproduced in the other panels, for reference, with grey
circles. Legends in each panel label the simulation series it
corresponds to.

Fig.~\ref{FigEtaRotEtaBarLCDM} is similar in concept to
Fig.~\ref{FigVfidVmaxLCDM}, but for the $\eta_{\rm rot}$-$\eta_{\rm bar}$
relation. Observed galaxies are shown in the top-left panel of
Fig.~\ref{FigEtaRotEtaBarLCDM} and are coloured by $V_{\rm max}$. We
shall use these two figures to discuss next the results of each
simulation series.

\subsubsection{APOSTLE}

As expected from the discussion above, APOSTLE galaxies (top-right panel in
Fig.~\ref{FigVfidVmaxLCDM}) show little diversity in their rotation
curve shapes, which track closely the loci of DMO LCDM halos (grey
line/shaded band in top-left panel). The upturn relative to the LCDM/NFW
line in massive galaxies results from the contribution of baryons in
the inner regions. Note that because of the relatively small
simulated volume there are few massive galaxies in APOSTLE. The
APOSTLE sample contains no slowly-rising, large core dwarf galaxies (i.e.,
systems well below the grey line), in disagreement with the
observational dataset considered here.

In terms of rotation curve shape vs baryon importance, we see from the
top-right panel of Fig.~\ref{FigEtaRotEtaBarLCDM} that few APOSTLE
galaxies are heavily dominated by baryons in the inner regions.  APOSTLE is
able to reproduce fairly well rapidly-rising rotation curves in
galaxies where baryons are unimportant (i.e., top-left corner of the
$\eta_{\rm rot}$-$\eta_{\rm bar}$ panel); these correspond to NFW-like
halos where the initial cusp has been, if anything, slightly
strengthened by the accumulation of baryons at the centre.

Slowly-rising rotation curves (i.e., $\eta_{\rm rot}\lsim 0.55$), as well as
heavily baryon-dominated galaxies ($\eta_{\rm bar}\gsim 0.4$) are not
present in these simulations. This comparison briefly summarizes the
known shortcomings of EAGLE/APOSTLE simulations to reproduce the
observed diversity of dwarf galaxy rotation curves \citep[e.g.,][]{Oman2015}.

\begin{figure*}
\includegraphics[width=0.7\linewidth]{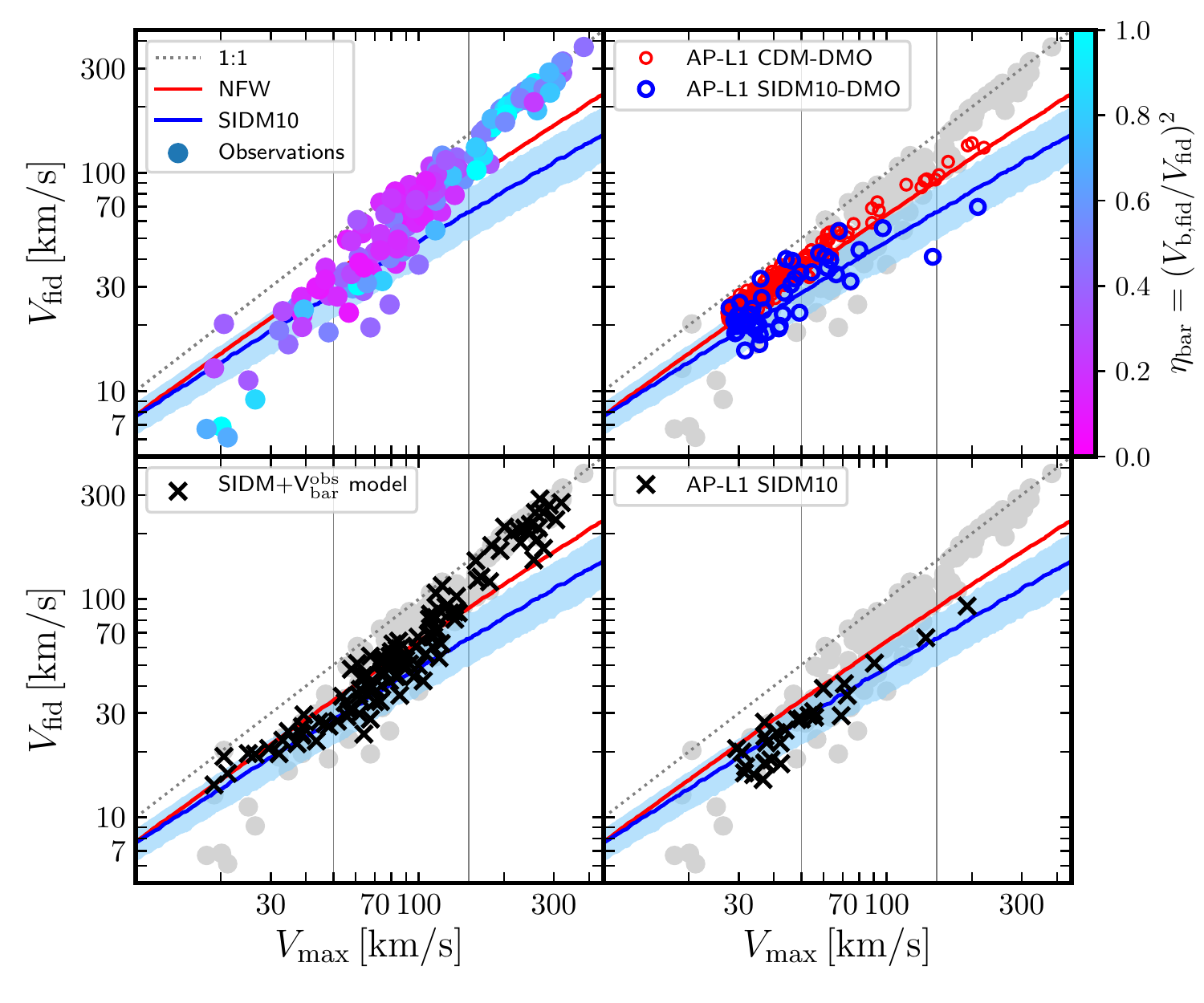}
\centering
\caption{As Fig.~\ref{FigVfidVmaxLCDM}, but for observations and SIDM. Top-left panel presents the
  observational sample, coloured by the inner baryon importance
  parameter, $\eta_{\rm bar}$. Other panels indicate the results of
  simulations, but also include observations, for
  reference, as grey circles. Legends in each panel identify the
  simulation series. See discussion in text.}
\label{FigVfidVmaxSIDM}
\end{figure*}

\subsubsection{NIHAO}
\label{SecNIHAODisc}

The comparison with NIHAO is shown in the bottom-left panels of
Figs.~\ref{FigVfidVmaxLCDM} and ~\ref{FigEtaRotEtaBarLCDM}. Starting
with Fig.~\ref{FigVfidVmaxLCDM}, NIHAO galaxies exhibit more
slowly-rising rotation curves than APOSTLE in the $V_{\rm max}$ range
$40$-$100$ km/s. This is the result of the cores created by baryonic
outflows in these runs. More massive galaxies have rapidly-rising
rotation curves, and thus no obvious cores, presumably because
baryonic outflows are less efficient in the deep potential wells of
these systems. Cores do not form in very low-mass galaxies either
(i.e., $V_{\rm max}<40$ km/s), in this case because too few stars form
in these systems to power the needed outflows
\citep[e.g.,][]{DiCintio2014a,DiCintio2014b}.

The cores in NIHAO help to reconcile LCDM with some of the
slowly-rising rotation curve systems that APOSTLE fails to
reproduce. Note, however, that because core formation is quite
efficient in the $40<V_{\rm max}/$km/s $<100$ range, NIHAO seems
unable to reproduce rapidly-rising rotation curves in that range
\citep[see also][]{Santos-Santos2018}.  Indeed, judging from the
lower-left panel of Fig.~\ref{FigVfidVmaxLCDM}, NIHAO's result do not
seem to capture the full diversity of dwarf galaxy rotation curves. In
addition, there is some evidence for cores in observed galaxies at the
lowest mass end (i.e., $V_{\rm max}<40$ km/s) again at odds with
NIHAO's results.

These shortcomings are also apparent in the bottom-left panel of
Fig.~\ref{FigEtaRotEtaBarLCDM}, where we see that NIHAO does not
reproduce the observed systems with ``cuspy'' rotation curves and
negligible baryon contribution (top-left corner), nor baryon-dominated
galaxies with large cores (bottom-right).

Regarding the latter (two examples of which, UGC05750 and IC2574, are
highlighted with circles in the bottom-left panel of
Fig.~\ref{FigEtaRotEtaBarLCDM}), \citet{Santos-Santos2018} have
suggested that the disagreement may have been caused by the assumption
of spherical symmetry when estimating circular velocities in
simulations \citep[see also][]{Dutton2019a}.

Indeed, these authors show in their Fig.~6 that the inner circular
velocities of two NIHAO galaxies (highlighted with squares in the
bottom-left panel of our Fig.~\ref{FigEtaRotEtaBarLCDM}) could be
substantially reduced by taking into account the actual flattened
geometry of the baryons when computing the gravitational
potential on the disc plane. This reduction could, in principle, lead to
lower values of $\eta_{\rm rot}$, as shown by the cyan dashed lines in
the bottom-left panel of Fig.~\ref{FigEtaRotEtaBarLCDM}.

However, because the NIHAO galaxies are actually dark matter dominated
at $r_{\rm fid}$, the reduction in $\eta_{\rm rot}$ is accompanied by
an even more substantial reduction in $\eta_{\rm bar}$, shifting the
galaxies to the bottom-left corner of the panel, rather than closer to
the observed galaxies. In other words, NIHAO galaxies may come close
to matching the {\it shape} of the rotation curves of UGC05750 and
IC2574, but are unable to reproduce, simultaneously, the importance of
baryons in their inner regions. This discussion highlights the power
of using {\it both} $\eta_{\rm rot}$ and $\eta_{\rm bar}$ as
diagnostics of the viability of a particular scenario meant to explain
the rotation curve diversity.

\subsubsection{EAGLE-CHT10}

Finally, we consider an alternative LCDM simulation run with the EAGLE
code, but where a higher star formation threshold allows baryonic
outflows to transform cusps into cores. The star formation and
feedback algorithm is quite different from NIHAO's so, in principle,
we do not expect the same results.  However, as may be seen in the
bottom-right panel of Fig.~\ref{FigVfidVmaxLCDM}, the results for
EAGLE-CHT10 are not too dissimilar to NIHAO's. There is, again, a
shortage of ``cuspy'' systems in the $70<V_{\rm max}/$km/s$<150$
range. We are unable to check for the presence of cores in
$V_{\rm max}<40$ km/s galaxies because of limited numerical
resolution. 

In terms of $\eta_{\rm rot}$ vs $\eta_{\rm bar}$
(Fig.~\ref{FigEtaRotEtaBarLCDM}), we see that, as in the case of
NIHAO, EAGLE-CHT10 has difficulty reproducing ``cuspy'' galaxies
where the baryon contribution is negligible. EAGLE-CHT10 fares a bit
better in terms of the largest cores, especially those with high
values of $\eta_{\rm bar}$, but the difference with NIHAO in this
respect is small.

\subsubsection{Summary}

It is clear that baryonic outflows can produce cores in dwarf
galaxies, reconciling in the process LCDM with systems with
slowly-rising rotation curves.  However, this mechanism, at least as
implemented in the NIHAO and EAGLE-CHT10 simulations we analyzed here,
is unable to account simultaneously for galaxies with ``cuspy''
rotation curves where the inner contribution of baryons is negligible.

The observed diversity seems to demand a mechanism that forms cores in
{\it some} galaxies only, while others retain (or re-form) a cusp,
independently of the baryonic mass contribution. This feature seems to
elude current simulations of this mechanism. It is unclear whether
this signals a fundamental shortcoming of the models, or just a need
to ``fine tune'' the numerical implementations. What is clear,
however, is that any successful explanation of the rotation curve
diversity should provide a natural explanation for the apparent
presence of cusps {\it and} cores in dwarfs and for their peculiar
relation to the importance of baryons in the inner regions.

\subsection{Diversity and SIDM}

Self-interacting dark matter is a distinct scenario for explaining the
rotation curve diversity, where cores in the inner dark matter density
profiles form not through baryonic outflows, but, rather, because of
the inward ``heat transfer'' driven by collisions
(``self-interactions'') between dark matter particles\footnote{ In
  principle, baryonic inflows and outflows may also affect the inner
  regions of SIDM halos, a complication that we shall ignore in this
  paper.}.  The simplest example of SIDM corresponds to elastic,
velocity-independent interactions where the magnitude of the effect is
controlled by a single parameter; the self-interacting cross section,
$s_{\rm si}$ \citep[see; e.g.,][]{Rocha2013}. This is, in principle, a
free parameter, but values between $0.1$ and $1$ cm$^2$/g lead to
tangible changes in the inner density profiles of dark matter halos
\citep[see; e.g., the recent review by][]{Tulin2018}.

An exploration of all SIDM alternatives is beyond the scope of this
paper, which adopts a single value of $s_{\rm si}=10$ cm$^2$/g. This is a rather
extreme value that, however, allows us to explore the maximal effect
of this mechanism without promoting a ``core collapse'' of the inner
regions \citep[e.g.,][]{Elbert2015}.  This model, which we shall refer
to hereafter as SIDM10, should be regarded as a limiting case where
cores are as large as this mechanism may be plausibly expected to yield.

We emphasize that more realistic SIDM models would include a
velocity-dependent cross section, which is needed to reduce the
effective cross section in galaxy clusters, where a cross section as
large as assumed here is clearly ruled out \citep[see; e.g.,][and
references therein]{Tulin2018}. Velocity dependence is indeed generic
with light force mediators, as would be expected for large cross
sections. We shall ignore these complications in our analysis, which
is not meant to rule in or out SIDM as a class, but rather to identify
further observational diagnostics useful for assessing the relative
performance of different scenarios.

\begin{figure*}
\includegraphics[width=0.7\linewidth]{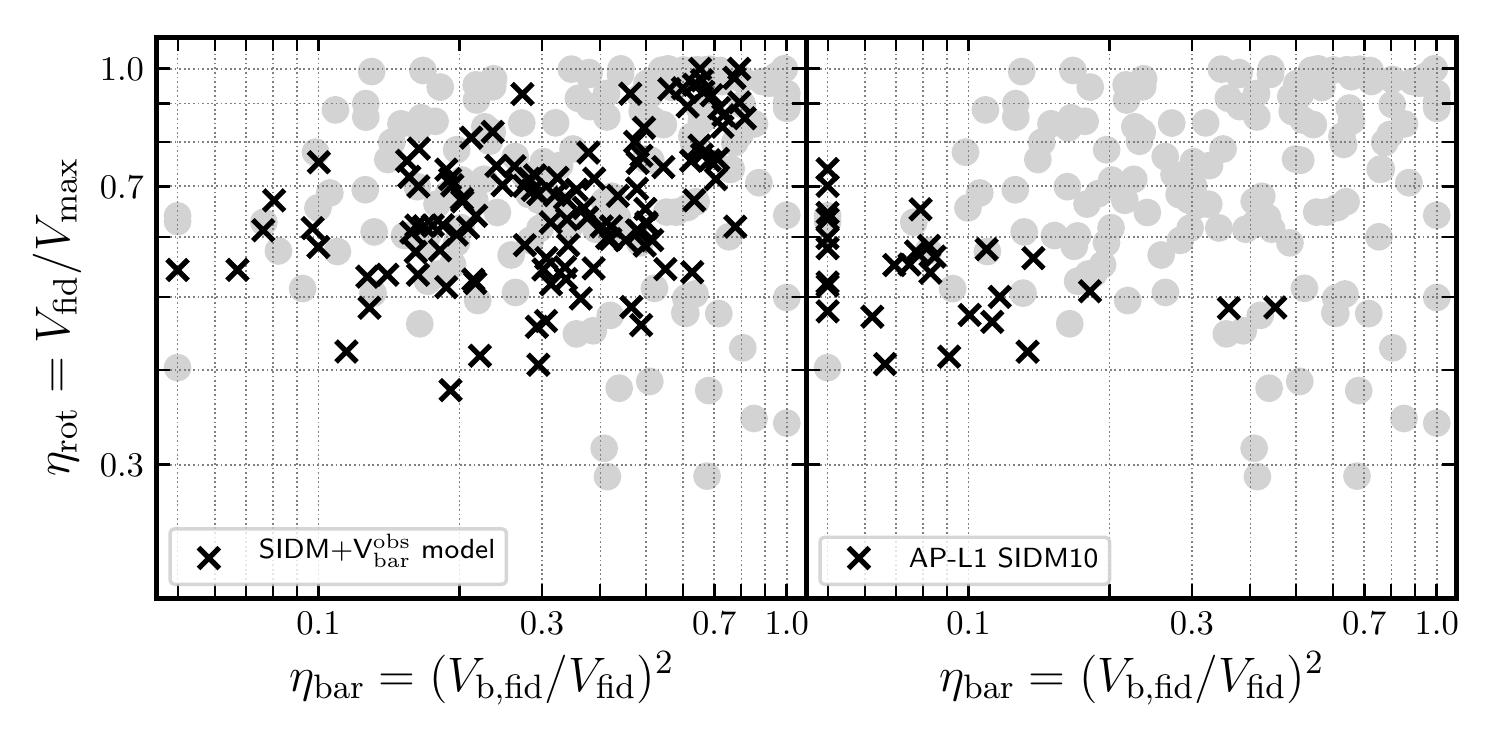}
\centering
\caption{Rotation curve shape parameter, $\eta_{\rm rot}$, vs baryonic
  importance parameter, $\eta_{\rm bar}$, for observations and SIDM
  simulations.  Left-hand panel shows observations (grey circles) and
  the results of the analytical model discussed in
  Sec.~\ref{SecSIDMmodel} (crosses). Right-hand panel corresponds to
  the SIDM10+baryon simulation discussed in
  Sec.~\ref{SecSIDMsim}. Systems with $\eta_{\rm bar}<0.05$ are shown
  at that value for clarity.  See discussion in text.  }
\label{FigEtaRotEtaBarSIDM}
\end{figure*}

\subsubsection{SIDM10: dark matter only}
\label{SecSIDMDMO}

We begin by exploring the effect of SIDM on the inner mass
distribution of dark matter halos in the dark-matter-only case. As
discussed above, interactions reduce the inner dark matter density and
promote the formation of a constant density core. This has little
effect on the maximum circular velocity of a halo, but can reduce
substantially the circular velocity at the inner fiducial radius.

We may see the resulting effect on the top-right panel of
Fig.~\ref{FigVfidVmaxSIDM}, where we plot the results of the DMO LCDM
AP-L1 runs (in red), together with results from the DMO SIDM10 run
(blue). Clearly, the values of $V_{\rm fid}$ at fixed $V_{\rm max}$
are substantially lower in the case of SIDM10 than for LCDM.

We can use the results of the AP-L1 and SIDM10 DMO runs to model the
DMO SIDM10 $V_{\rm fid}$-$V_{\rm max}$ relation, as well as its
scatter, starting from the LCDM $M_{200}$-concentration relation
\citep[and its scatter;][]{Ludlow2016}. Details of the procedure are
given in an Appendix (Secs.~\ref{SecSIDMCDMEmpir} and ~\ref{SecSIDMCDMAnMod}), but we describe it briefly below, for
completeness.

The inner mass profile of an SIDM halo is well approximated by a
non-singular isothermal sphere, which is fully described by a pair of
parameters \citep[see; e.g., p.228 of][]{BinneyTremaine1987}. These may be taken to be the
central density, $\rho_0$, and a scale radius, $r_{\rm 0}$, or,
alternatively, a velocity dispersion, $\sigma_0$. For SIDM10, these parameters correlate closely with the
corresponding LCDM parameters: for example, there is a close
relation between the $V_{\rm max}$ of an LCDM halo and the
characteristic $\sigma_0$ of the counterpart SIDM halo that forms when the
{\it same} initial conditions are evolved with self-interactions turned on.

Likewise, there is a strong correlation between $r_{\rm 0}$ and the
characteristic radial scale of an LCDM halo, best expressed through
$r_1$, the radius where, on average, one interaction per particle is
expected per Hubble time. These correlations are sensitive to the
value of $s_{\rm si}$ adopted; we show them for $s_{\rm si}=10$
cm$^2$/g in Fig.~\ref{FigCDMSIDMScatter} for matching halo pairs
identified in LCDM and SIDM DMO simulations of one of the AP-L1
volumes.

These same correlations may be used to generate a population of SIDM10
halos that include both the original scatter in the LCDM
mass-concentration relation, but also the scatter in the relations
that link LCDM and SIDM10. The results of this procedure, in terms of
$V_{\rm fid}$ and $V_{\rm max}$ are shown by the blue thick line (and
shaded area, which delineate the $10$:$90$ percentiles) in the
top-right panel of Fig.~\ref{FigVfidVmaxSIDM}. To make further
progress we need to model the effect of baryons into this population
of SIDM10 halos, either through modeling or direct simulation. We
pursue this below.

\subsubsection{SIDM10 + baryons: a model}
\label{SecSIDMmodel}

As baryons accumulate at the centre of an SIDM halo, they are expected
to deepen the central gravitational potential. This should cause the
surrounding dark matter to respond by contracting and, for large
enough perturbations, by rebuilding the inner cusp. This process has
been explored using analytical techniques and simulations of isolated
systems in prior work
\citep{Kaplinghat2014,Kamada2017,Ren2019,Creasey2017}, which argue
that the effect should be strong enough in practice to produce cuspy
and cored rotation curves, and may account for the observed rotation
curve diversity.

In this scenario, slowly-rising rotation curves reflect systems where
self-interactions have carved a core, and where the baryons are not
important enough to rebuild the cusp. At the other extreme,
rapidly-rising rotation curves should generally correspond to systems
where baryons deepened the central potential and are gravitationally
important enough to rebuild the central dark matter cusp. These two
features are, at first glance, at odds with the trends for dwarf
galaxies highlighted in Fig.~\ref{FigEtaRotEtaBar}, where it is clear
that there are many ``cuspy'' systems where baryons are unimportant
and, in addition, that baryons actually do play an important role at
the centre of systems with the largest cores. Can SIDM models resolve
this apparent disagreement?

We explore this by using an analytical model, described in detail in
the Appendix (Sec.~\ref{SecSIDMCDMIncBar}), to estimate the response of SIDM halos to the
accumulation of baryons at the centre, and to compare the resulting
rotation curves with observations. In particular, we use the {\it
  actual} baryonic mass profile of individual galaxies and place them
in randomly selected SIDM halos of the appropriate $V_{\rm max}$ to
verify whether the resulting rotation curves reproduce the observed
diversity.

Our modeling proceeds as follows. For each observed galaxy in our
sample, we choose a random SIDM10 halo of matching $V_{\rm max}$
(Sec.~\ref{SecSIDMDMO}) and compute the change in the inner dark
matter distribution expected from the addition of the baryons. The
procedure uses the full baryonic distribution of each galaxy (given by
$V_{\rm bar}(r)$) to compute the response of the dark matter (see
details in Sec.~\ref{SecSIDMCDMIncBar}), and results in a new rotation curve for the
galaxy. The procedure preserves $V_{\rm max}$ and $V_{\rm bar}(r)$ of
each galaxy, but modifies its rotation speed at the inner fiducial
radius, $V_{\rm fid}$.

The results of this modeling are presented in the bottom-left panel of
Fig.~\ref{FigVfidVmaxSIDM}. At first glance, this results in a wider
range of rotation curve shapes compared with the LCDM simulations shown in
Fig.~\ref{FigVfidVmaxLCDM}. This is, however, partly a result of the
wider diversity of baryonic profiles probed here, which matches, by
construction, exactly those of the observed sample.

In particular, the SIDM10 model successfully reproduces the steeply
rising rotation curves of massive galaxies (i.e., $V_{\rm max}>150$
km/s), despite the fairly large cores imposed by SIDM. In lower mass
systems the diversity is reproduced less well, with fewer systems near
the $1$:$1$ line and, despite the extreme value of $s_{\rm si}$
adopted, few cores as large as observed. This is especially true at
the very low mass end, where SIDM cores seem too small to affect the
rotation curve at $r_{\rm fid}$.

This conclusion is supported by the distribution of galaxies in the
$\eta_{\rm rot}$-$\eta_{\rm bar}$ plane (left-hand panel of
Fig.~\ref{FigEtaRotEtaBarSIDM}) where it is clear that there are few
galaxies in the upper-left and bottom-right corners of this plot. Like
the BICC models discussed in Sec.~\ref{SecDivLCDM}, SIDM10 has
difficulty accounting for both the observed population of
sharply-rising rotation curves without a dominant inner baryonic
contribution (top-left corner), and for the very slowly-rising
rotation curves where baryons play an important role near the centre
(bottom-right corner).  On its own, the SIDM hypothesis seems thus
unable to explain fully the observed diversity of rotation curve shapes.

Note that our analysis addresses whether observed galaxies, if placed
in SIDM halos of randomly sampled concentration (and not particular halos chosen to fit the
observed rotation curves), would exhibit the observed diversity. This
differs from earlier work on this topic, which addressed whether it
is {\it possible} to obtain adequate fits to individual rotation
curves in SIDM halos by allowing the size of the core (which is
intimately linked to the assumed concentration) to vary as a
free parameter \citep[see; e.g.,][and references
therein]{Ren2019}. Such work reproduces the diversity {\it by
  construction}, by carefully choosing concentration values for the halos
of individual galaxies so as to fit their rotation curves. The
distribution of required concentrations, however, is quite different from what would
be expected from randomly sampling the halo mass-concentration
relation \citep[see][for a similar
argument]{Creasey2017}..

\begin{figure}
\includegraphics[width=\linewidth]{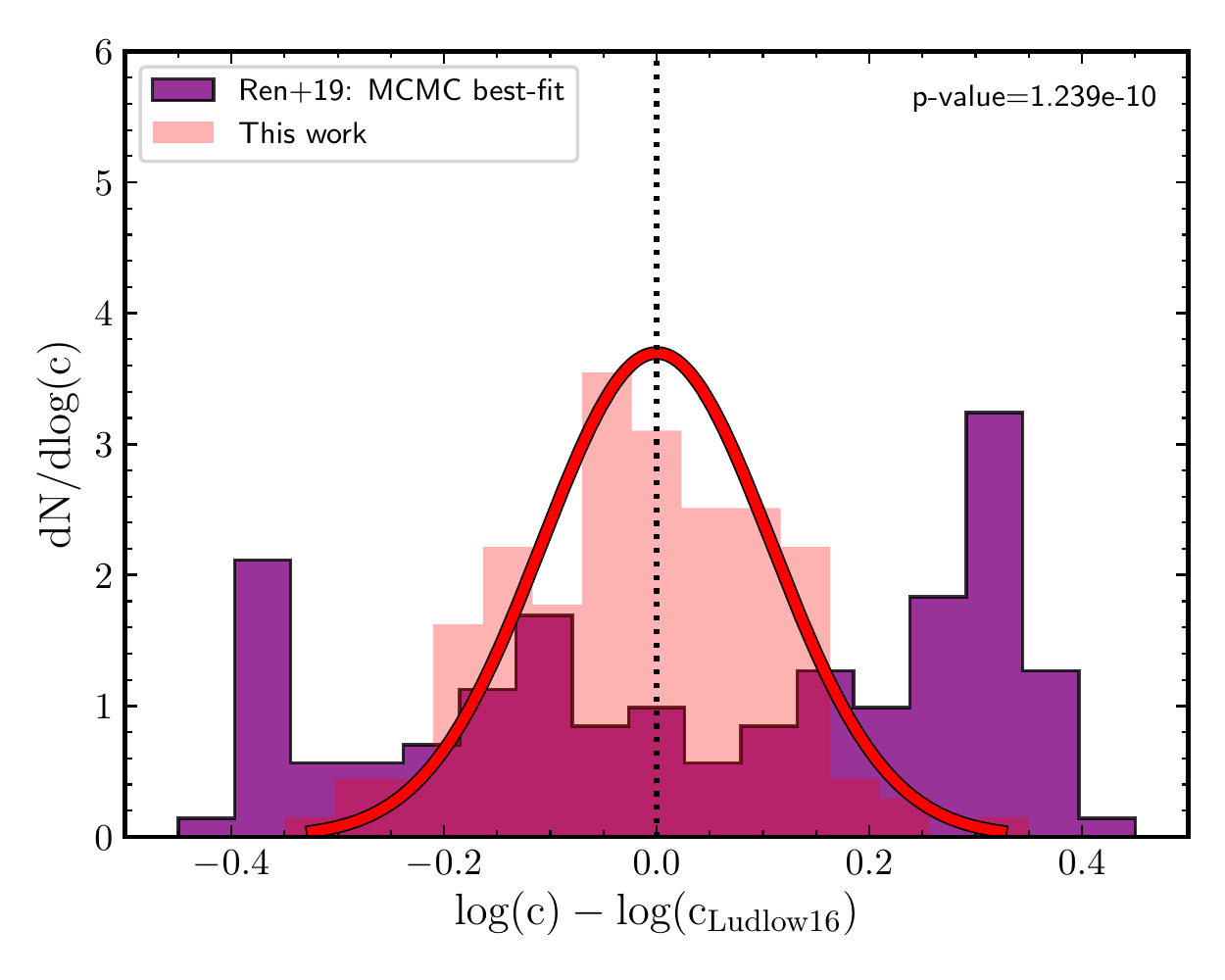}
\centering
\caption{Distribution of halo concentrations about the mean expected
  in LCDM for halos of given $V_{\rm max}$ (red solid curve; from
  \citet{Ludlow2016}'s $M_{200}$(c) relation plus scatter). The concentrations adopted in
  the SIDM model used in this work is shown by  the pink histogram; those used in the SIDM
  model of \citet{Ren2019} are shown by the purple histogram. See text
  for further details.}
\label{FigConc}
\end{figure}

We illustrate this in Fig.~\ref{FigConc}, where we contrast the
probability distribution of concentrations adopted by \citet{Ren2019}
(i.e., their ``MCMC best-fit''model; see their Table S2) with the
concentration distribution used in our work (pink
histogram). Clearly, the two distributions are substantially and
significantly different.

In other words, fitting the observed rotation curve diversity with
SIDM requires carefully chosen concentrations which, taken as an
ensemble, are quite unlike what is expected from
cosmological simulations (Gaussian curve in Fig.~\ref{FigConc}).
While it remains {\it possible} to match rotation curves with SIDM,
the improbable distribution of required concentrations detracts, in
our opinion, from the appeal of the SIDM scenario as an
explanation of the diversity of dwarf galaxy rotation curves.

\subsubsection{SIDM10 + baryons: a simulation}
\label{SecSIDMsim}

The results of a cosmological hydrodynamical simulation that includes
a self-interaction cross section of $s_{\rm si}=10$ cm$^2$/g and the
EAGLE/APOSTLE star formation/feedback implementation are presented in
the bottom-right panel of Fig.~\ref{FigVfidVmaxSIDM}. (Details of the
simulation are given in Sec.~\ref{SecSims}.) Recall that, because of
the low star formation density threshold adopted in this model, we do
not expect baryonic outflows to effect large changes in the inner dark
matter density profile. Indeed, the simulation results show that, in
terms of $V_{\rm fid}$-$V_{\rm max}$, the simulated galaxies follow
approximately the dark matter-only results (indicated by the blue
shaded area). As for the analytic model discussed in
the previous subsection, low-mass simulated galaxies fail to populate
the rapidly-rising ($1$:$1$) rotation curve regime and to account for
the largest observed cores.

These conclusions are again supported by inspection of the
$\eta_{\rm rot}$-$\eta_{\rm bar}$ plane, shown in the right-hand panel
of Fig.~\ref{FigEtaRotEtaBarSIDM}. There is again a clear dearth of
``cuspy'' systems where the baryons are gravitationally unimportant
(upper-left corner), and of ``cored'' systems where baryons dominate
the inner regions (bottom-right corner).  Indeed, baryons do not seem
to play an important role in any simulated SIDM10 galaxy, although this is
likely a result of the particular star formation algorithm adopted in
the EAGLE/APOSTLE code. These conclusions echo those of the previous
subsection, and highlight the difficulty faced by models like SIDM, where most
halos develop cores, to account naturally for the observed rotation curve diversity.

\subsection{Diversity and non-circular motions }
\label{SecNonCirc}

We consider next the effects that non-circular motions may have on
the interpretation of the rotation curve diversity. As discussed by
\citet{Oman2019} and \citet{Marasco2018}, the triaxial potential of
LCDM halos may induce non-circular motions in the gaseous discs of
dwarf galaxies. In the simplest case, closed orbits in the disc plane
become elliptical and the azimuthal speed of the gas varies along the orbit, from
maxima along the orbital minor axis (pericentres) of the ellipse to
minima along the orbital major axis (apocentres). This may lead to
different rotation curves for the {\it same} galaxy, depending on how
the line of nodes of a particular sky projection aligns with the
orbital principal axis \citep[see also][]{Hayashi2006}. In particular, large underestimates of
the inner circular velocity may result when the projected kinematic major axis
is aligned with the orbital apocentres, mimicking the effect of a
core.

Non-circular motions may affect the inferred values of both
$V_{\rm fid}$ and $V_{\rm max}$, introducing scatter in the tight
relation between these two parameters expected in LCDM simulations
where cores do not form, such as APOSTLE. We show the effect by
selecting two dwarf galaxies from the AP-L1 sample and projecting them
along 24 different lines of sight at fixed inclination,
$i=60^o$. Synthetic HI data cubes are constructed for each projection,
which are then modelled using $^{\rm 3D}$BAROLO, a publicly available
tilted-ring processing tool \citep{DiTeodoro2015,Iorio2017}. Details
of the modeling as applied to the APOSTLE galaxies may be found in
\citet{Oman2019}.

\begin{figure}
\includegraphics[width=\linewidth]{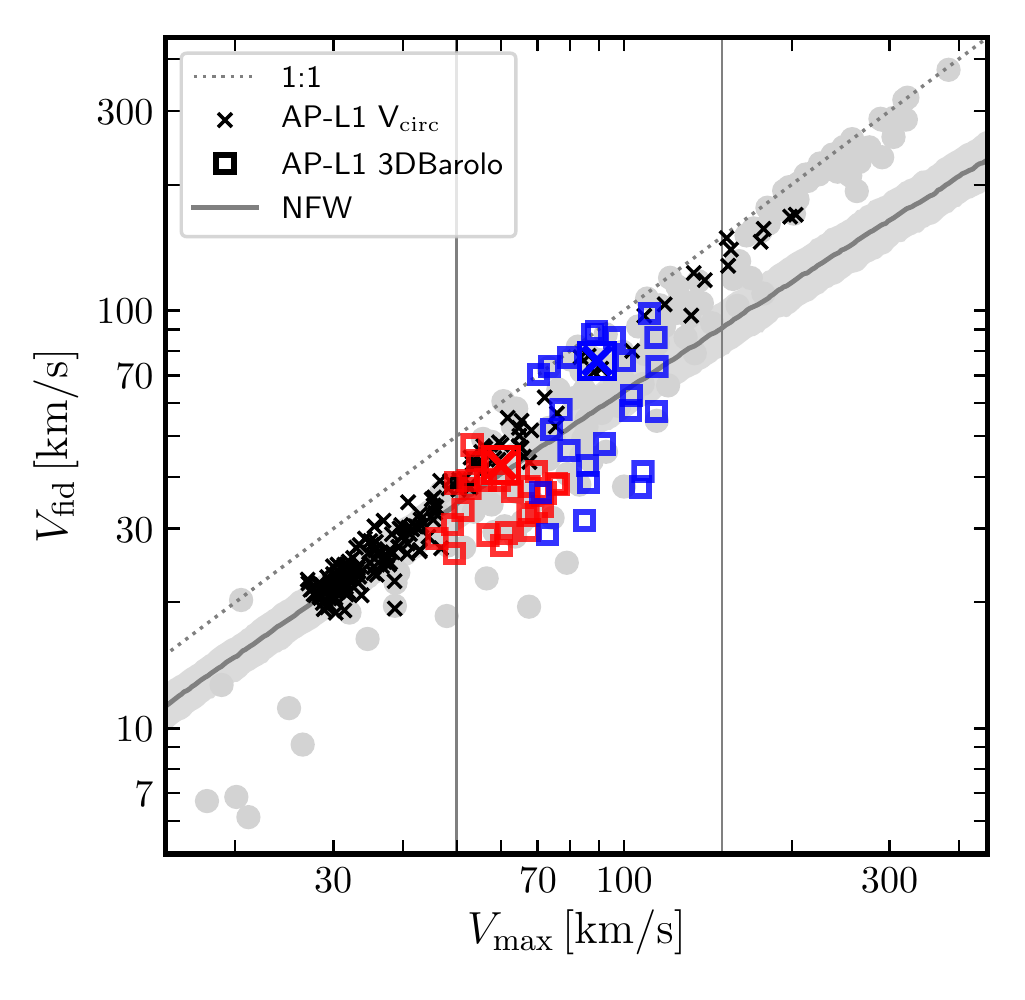}
\includegraphics[width=\linewidth]{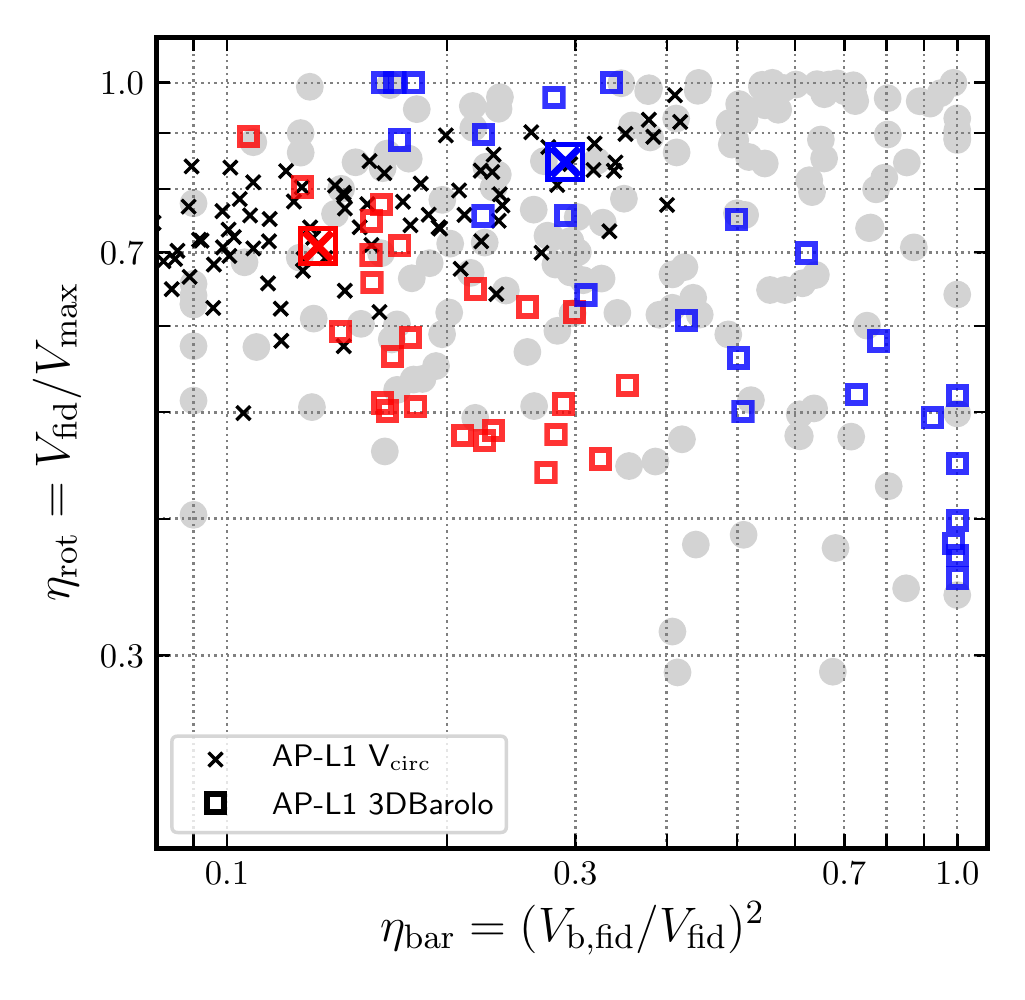}
\centering
\caption{{\it Top:} Same as the top-right panel of
  Fig.~\ref{FigVfidVmaxLCDM}, but including the results of mock HI
  observations of two simulated galaxies from APOSTLE, analyzed with
  $^{\rm 3D}$BAROLO \citep{DiTeodoro2015,Oman2019}.  AP-L1-V6-5-0 is
  shown in blue, AP-L1-V4-14-0 in red.  The results of $24$ different
  lines of sight at fixed inclination ($i=60^o$) are shown, with
  different colours for each of the two galaxies. {\it Bottom:} Same as
  the top-right panel of Fig.~\ref{FigEtaRotEtaBarLCDM}, for the same
  two galaxies as in top panel. The large boxes with a cross indicate
  the values obtained from the circular velocity curve rather than
  from $^{\rm 3D}$BAROLO. }
\label{FigNonCirc}
\end{figure}

Each of the two galaxies is shown with different colours in
Fig.~\ref{FigNonCirc}, illustrating the dramatic impact that
non-circular motions may have on the $V_{\rm fid}$-$V_{\rm max}$ and
$\eta_{\rm rot}$-$\eta_{\rm bar}$ correlations. At least for these two
galaxies the effects of non-circular motions are quite substantial. For
galaxy AP-L1-V6-5-0 \citep[shown in blue; notation as in][]{Oman2019},
the maximum circular velocity may vary from $70$ km/s to slightly over
$100$ km/s and the inferred rotation speed at the inner fiducial
radius from $\sim 30$ to $100$ km/s, depending on
projection. Non-circular motions of this magnitude can clearly explain
much of the observed diversity in the $V_{\rm fid}$-$V_{\rm max}$
plane.

In particular, the {\it same} galaxy could end up being classified
either as a rapidly-rising rotation curve where baryons are relatively
unimportant (top-left corner in the bottom panel of
Fig.~\ref{FigNonCirc}), or as a galaxy with a large core where baryons
dominate the inner regions (bottom-right corner of the same
panel). This is because non-circular motions tend to scatter galaxies
along the same band in the $\eta_{\rm rot}$-$\eta_{\rm bar}$ plane
traced by dwarf galaxies (see galaxies with $V_{\rm max}<150$ km/s in
Fig.~\ref{FigEtaRotEtaBar}). Non-circular motions thus provide an
appealing explanation for this puzzling trend, which, as we have
discussed, is not reproduced well in other scenarios.

Although attractive, the non-circular motion explanation of the
diversity also suffers from shortcomings. In particular, it may
be argued that the magnitude of non-circular motions required to
spread galaxies across the whole $\eta_{\rm rot}$-$\eta_{\rm bar}$
plane is quite substantial, whereas many observed galaxies show
quieter velocity fields, without clear obvious distortions
\citep[][]{Trachternach2008}. However, \citet{Oman2019}
report that at least two galaxies with obvious ``cores'' (DDO 47 and
DDO 87) show clear signs of non-circular motions and emphasize that
these may be difficult to detect because of the large number of free
parameters involved in tilted-ring model fits.

A second difficulty refers to our earlier discussion of
Fig.~\ref{FigEtaRotSB}, which shows that ``cores'' are only manifest in
low surface brightness/density galaxies. If diversity were due mostly
to non-circular motions, why would they only affect low surface
brightness galaxies? There is no clear answer to this, but one possibility is that in galaxies with
highly-concentrated baryonic components the halo responds by
becoming more spherical, reducing the importance of non-circular
motions \citep[e.g.,][and references therein]{Abadi2010}. For the same
reason, we emphasize that non-circular motions of the kind described
here would only be expected in LCDM halos that have preserved their
original cusp and triaxiality, since mechanisms that
erase the cusp are also likely to sphericalize the halo. This is a
well known feature of SIDM \citep[e.g.,][]{Miralda-Escude2002} that is
likely to apply as well to baryon-induced cores.

A detailed re-analysis of all ``cored'' galaxies designed specifically
to test the non circular-motion hypothesis, together with a concerted
effort to infer rotation curves from synthetic 2D velocity fields for
more simulated galaxies is clearly needed in order to assess the true
viability of this scenario. However, at least from the evidence
presented in Fig. ~\ref{FigNonCirc}, it would be premature to rule out
non-circular motions as one of the driving causes of the observed
diversity.

\begin{figure}
  \includegraphics[width=\linewidth]{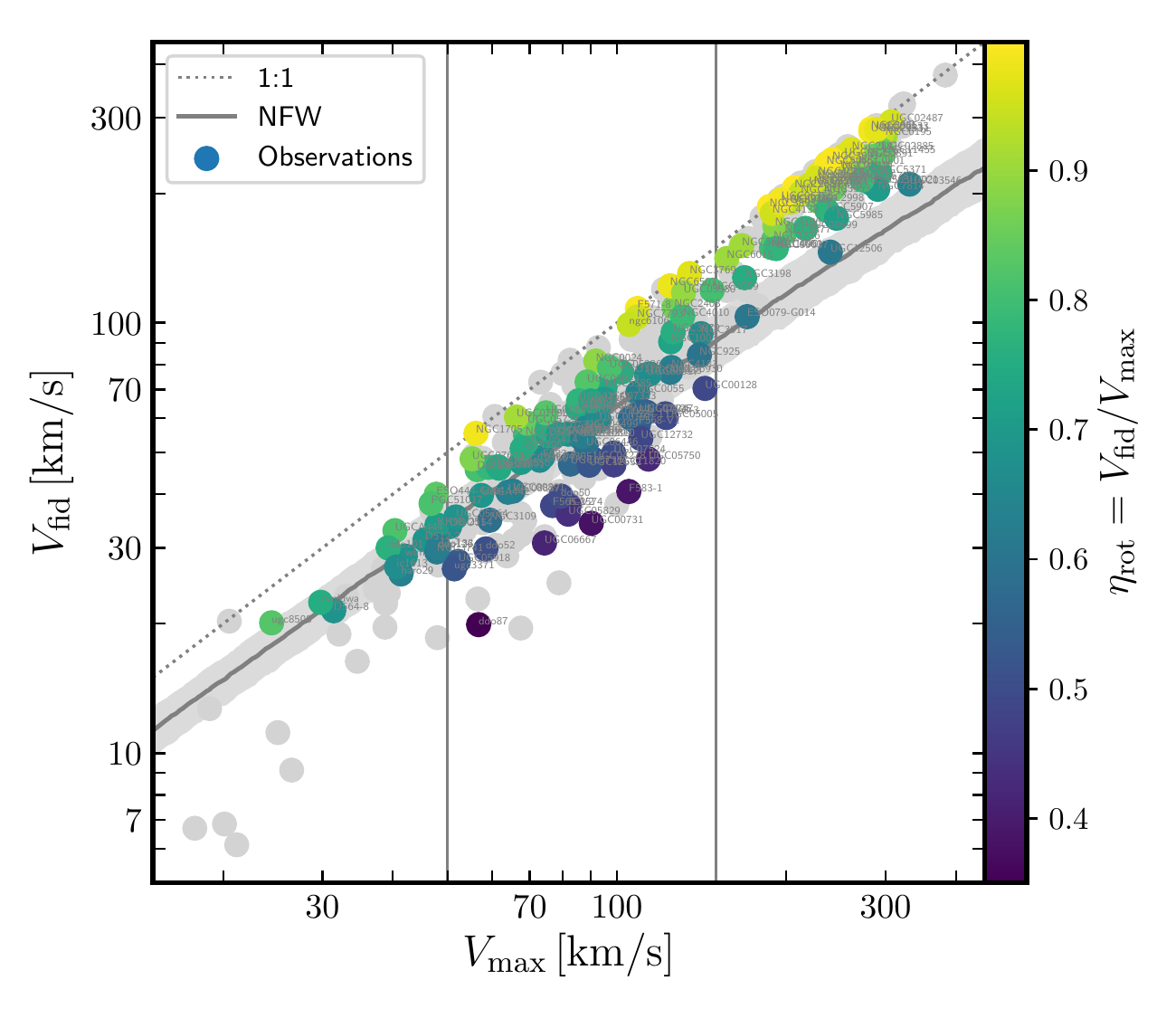}
  \includegraphics[width=\linewidth]{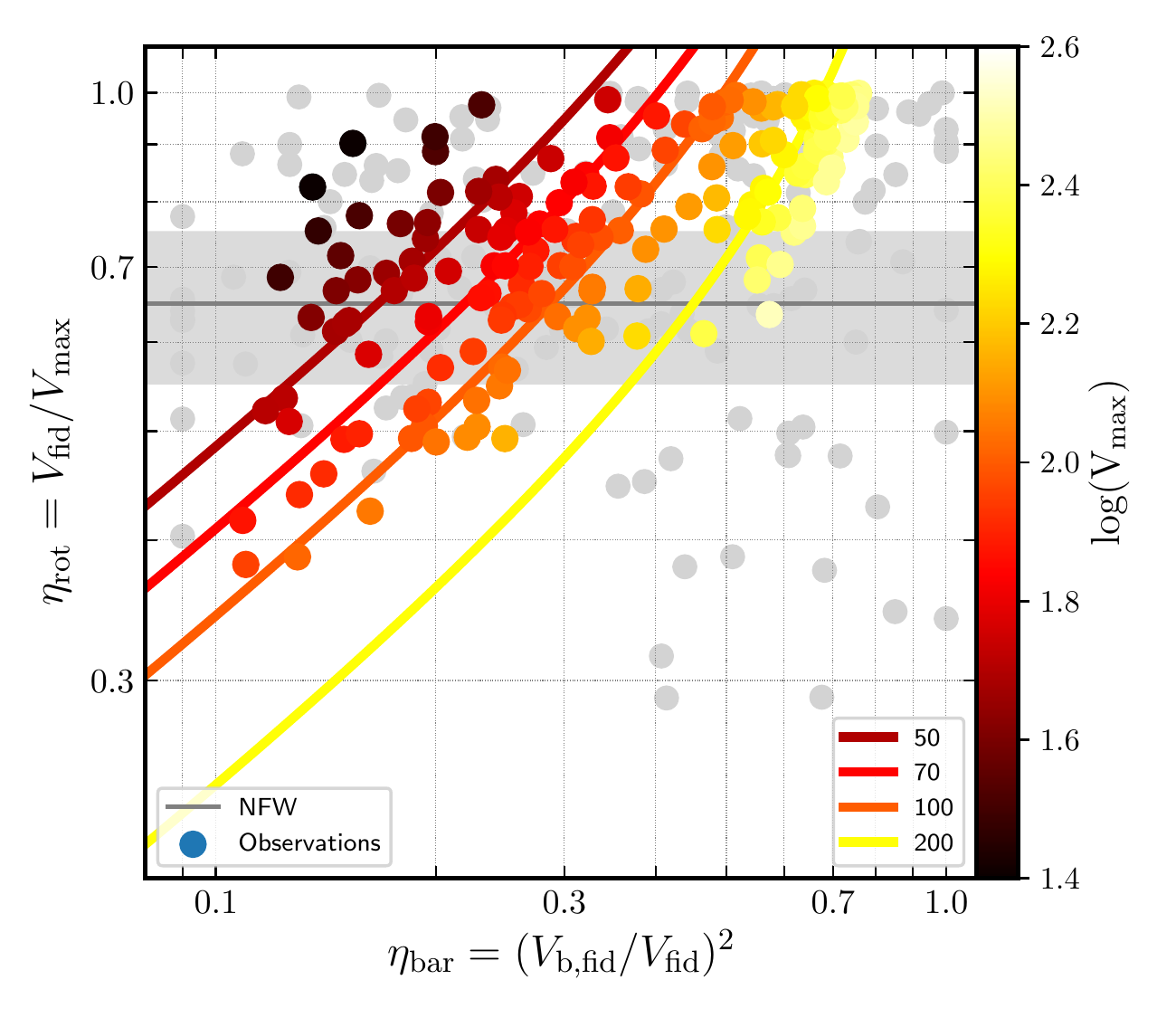}
  \centering
  \caption{{\it Top:} Same as 
  Fig.~\ref{FigVfidVmax}, but with the original data shown in
  grey. The coloured circles correspond to re-deriving $V_{\rm fid}$
  and $V_{\rm max}$ for each galaxy using solely the $V_{\rm bar}(r)$
  profile and the MDAR.  {\it Bottom:} Same as
Fig.~\ref{FigEtaRotEtaBar}, with the original data in grey. Coloured
circles correspond to re-deriving $\eta_{\rm rot}$
  and $\eta_{\rm bar}$ using only the $V_{\rm bar}(r)$
  profile and the MDAR. The MDAR forces galaxies of given $V_{\rm
    max}$ to lie along curves such as those shown in this panel, for a
few select values of $V_{\rm max}$.}
\label{FigVEMDAR}
\end{figure}

\begin{figure}
\includegraphics[width=\linewidth]{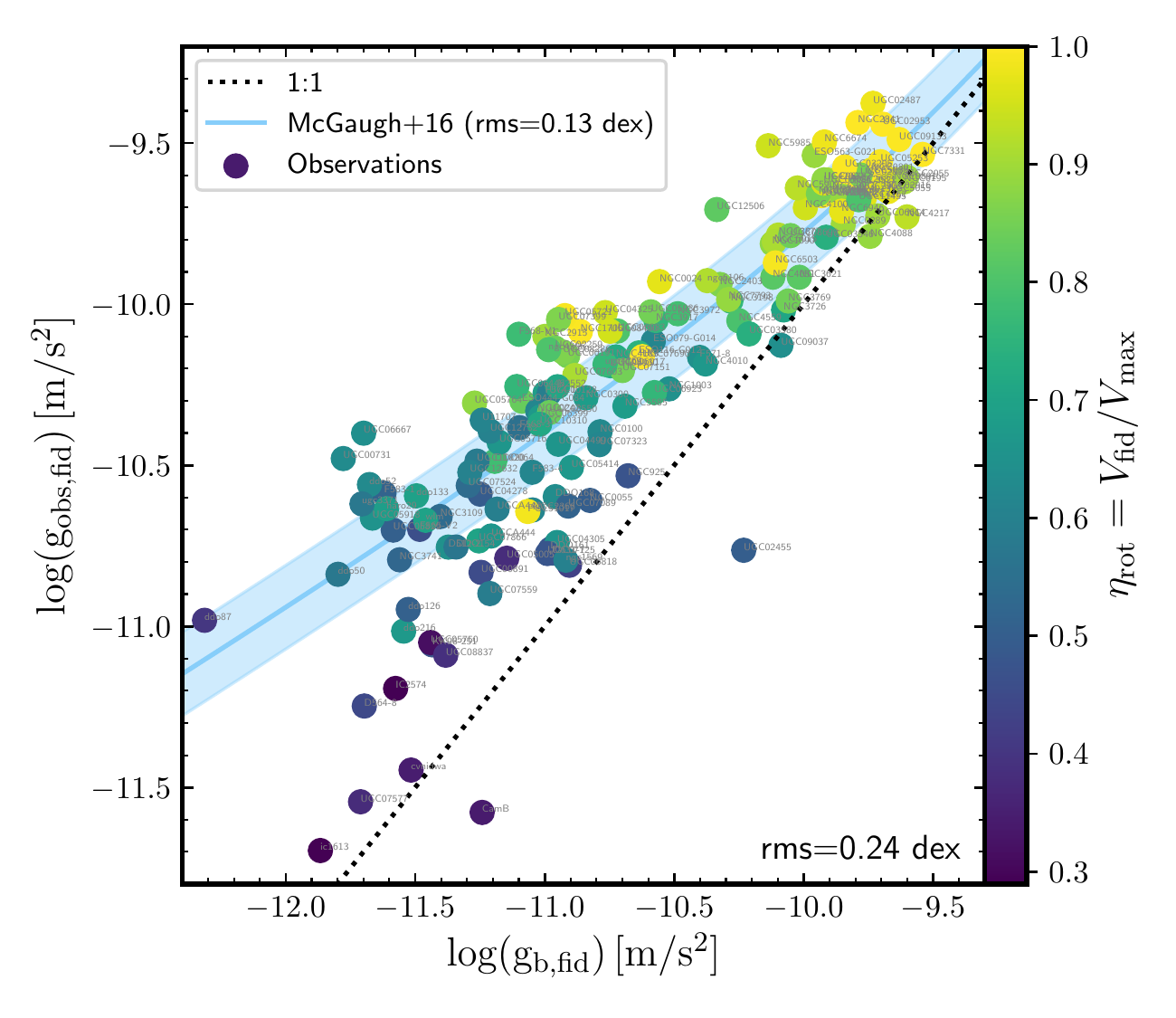}
\centering
\caption{Observed radial acceleration at the inner fiducial radius,
    $g_{\rm obs,fid}=V_{\rm fid}^2/r_{\rm fid}$,
   vs the
  baryonic acceleration at the same radius,
    $g_{\rm b,fid}=V_{\rm b,fid}^2/r_{\rm fid}$
   for all
  galaxies in our observational sample. Each galaxy is coloured by the
  rotation curve shape parameter, $\eta_{\rm rot}$. Note that
  ``cored'' galaxies with slowly-rising rotation curves (i.e.,
  $\eta_{\rm rot}\lsim 0.55$) typically fall outside of the MDAR relation
  proposed by \citet{McGaugh2016}, shown by the blue shaded area. The
  diversity in observed rotation curves is thus not caused by the MDAR
  through variations in the gravitational acceleration at
  $r_{\rm fid}$ induced by the baryonic component.  }
\label{FigMDAR}
\end{figure}

\subsection{Diversity and MDAR}\label{SecMDAR}

Finally, we explore whether the observed rotation curve shape
diversity could be a consequence of the ``mass
discrepancy-acceleration relation'' (MDAR) proposed by
\citet{McGaugh2016}. The MDAR posits that, at all radii, and for all
galaxies, observed accelerations, $g_{\rm obs}(r)=V_{\rm rot}^2(r)/r$
are uniquely determined by the accelerations expected from
the baryonic distribution, $g_{\rm bar}(r)=V_{\rm bar}^2(r)/r$. This,
of course, implies a unique relation between $V_{\rm bar}$ and $V_{\rm
  rot}$ at all radii, so that the full rotation curve may be derived solely
from the distribution of baryons.

In this scenario, the diversity of rotation curve shapes must
therefore follow from variations in the baryonic mass distribution,
which may in principle induce a large spread of inner acceleration
values in galaxies of given $V_{\rm max}$.  This issue has been
studied by \citet{Ghari2019}, who considered the original SPARC
dataset as well as revisions to the SPARC dataset proposed by
\citet{Li2018}\footnote{Assuming that the MDAR is indeed valid, these
  authors adjust the distances, inclinations, and mass-to-light ratios
  of SPARC galaxies so as to derive a much tighter relation than
  originally presented by \citet{McGaugh2016}. With these adjustments
  the overal scatter in the MDAR may be reduced from $\sim 0.130$ dex
  to $\sim 0.057$ dex.}, and concluded that, if the MDAR does indeed
hold, a sizable scatter in $V_{\rm fid}$ is indeed expected, mainly
driven by the structural diversity of galaxies of given $V_{\rm max}$.

We show this in the top panel of Fig.~\ref{FigVEMDAR}, where the
coloured circles correspond to re-deriving the values of $V_{\rm fid}$
and $V_{\rm max}$ for each galaxy using only the $V_{\rm bar}(r)$
profile and the MDAR (assuming negligible scatter). The grey circles
in the same figure are the original data, as presented in
Fig.~\ref{FigVfidVmax}. Note that although the main trends remain, the
diversity in the dwarf galaxy regime has been reduced, and most dwarfs
now lie closer to the LCDM prediction, indicated by the ``NFW'' band
in the same figure. We have explicitly checked that
using the \citet{Li2018} revisions to $V_{\rm bar}(r)$ for SPARC
galaxies does not change appreciably this conclusion; if anything, it
reduces the diversity even further.

In the context of our discussion, we note that the MDAR is
incompatible with the trends between inner baryon dominance and
rotation curve shape ($\eta_{\rm rot}$ vs $\eta_{\rm bar}$)
highlighted when discussing dwarf galaxies in Fig.~\ref{FigEtaRotEtaBar}. Indeed, it is
straightforward to show that the MDAR links these two parameters so
that galaxies must follow curves in the $\eta_{\rm
  rot}$-$\eta_{\rm bar}$ plane like those shown in the bottom panel of
Fig.~\ref{FigVEMDAR}, depending solely on their $V_{\rm max}$. Dwarf
galaxies, in particular, are constrained to the top-left and are
excluded from the bottom-right region of this panel. In other words,
dwarf galaxies with slowly-rising rotation curves (``cores'')  and where baryons
dominate the inner regions should not exist in the MDAR scenario.

This is illustrated by the coloured circles in
the bottom panel of Fig.~\ref{FigVEMDAR}, which correspond to assuming that the
MDAR holds (with negligible scatter), so that only $V_{\rm bar}(r)$ is
used to compute $\eta_{\rm rot}$ and $\eta_{\rm bar}$ for each
observed galaxy. As is clear, the trend from top-left to bottom-right
drawn by dwarf galaxies in Fig.~\ref{FigEtaRotEtaBar} is not present
anymore, and no dwarfs are found at the bottom-right region of the
plot. The aforementioned trend must therefore arise, in the MDAR
interpretation, as a consequence of (substantial) errors in the
 rotation velocities.

This is reminiscent of the discussion in the preceding subsection
(Sec.~\ref{SecNonCirc}), where we argued that errors in the observed
rotation speeds (due, in that case, to non-circular motions) may
reconcile the LCDM AP-L1 results with the observed diversity.  Indeed,
note that there is a strong similarity between dwarf galaxies in
Fig.~\ref{FigVEMDAR} and LCDM galaxies in the AP-L1 runs
(see; e.g., the crosses in the bottom panel of Fig.~\ref{FigNonCirc}):
in both cases dwarfs cluster at the upper-left corner of this
figure. This similarity has already been hinted at by
\citet{Navarro2017} and \citet{Ludlow2017}, who argued that the MDAR relation is
actually readily reproduced in LCDM: since dwarf galaxies with large
``cores'' are unexpected in LCDM, they should also
deviate systematically from the MDAR, at least in their inner
regions. 

We may see this directly in Fig.~\ref{FigMDAR}, where we show how
closely the total and baryonic accelerations correlate at the inner
fiducial radius for all galaxies in our sample. The MDAR relation from
\citet{McGaugh2016}, including its quoted scatter, is shown by the
light blue shaded region. Observed galaxies are coloured by the shape
parameter, $\eta_{\rm rot}$, and are seen to follow only approximately
the MDAR relation at $r_{\rm fid}$. Indeed, the rms scatter in
$g_{\rm obs}$ is $0.24$ dex, substantially larger than the $0.13$ dex
obtained by \citet{McGaugh2016} by combining data for all galaxies and
radii. In addition, most galaxies with large ``cores'' (i.e.
$\eta_{\rm rot}<0.5$) fall systematically off and below the MDAR
relation \citep[see also][ who report a similar
result]{SantosSantos2016,Ren2019}.

In conclusion, although variations in the baryonic distribution of
galaxies may lead to some diversity in rotation curve shapes, the
subtler trends between such shapes and the importance of baryons in
the inner regions implied by the data are incompatible with the
MDAR. On the other hand, if the MDAR actually holds, then the baryonic
trends must result from errors in the data. Since the same data was
actually used to derive the MDAR in the first place, this reduces, in
our opinion, the appeal of the MDAR as a possible explanation for the
diversity of dwarf galaxy rotation curve shapes.

\section{Summary and Conclusions}
\label{SecConc}

Dwarf galaxy rotation curves are challenging to reproduce in the
standard Lambda Cold Dark Matter (LCDM) cosmogony. In some galaxies,
rotation speeds rise rapidly to their maximum value, consistent with
the circular velocity curves expected of cuspy LCDM halos. In others,
however, rotation speeds rise more slowly, revealing large ``inner
mass deficits'' or ``cores'' when compared with LCDM halos
\citep[e.g.,][]{DeBlok2010}. This diversity is unexpected in LCDM, where,
in the absence of modifications by baryons, circular velocity curves
are expected to be simple, self-similar functions of the total halo
mass \citep{Navarro1996a,Navarro1997,Oman2015}. We examine in this
paper the viability of different scenarios proposed to explain the
diversity, and, in particular, the apparent presence of both cusps and
cores in dwarfs.

A 
first scenario (BICC; ``baryon-induced cores/cusps'') envisions the
diversity as caused by the effect of baryonic inflows and outflows
during the formation of the galaxy, which may rearrange the inner dark
matter profiles: cores are created by baryonic blowouts but cusps can
be recreated by further baryonic infall \citep[see;
e.g.,][]{Navarro1996b,Pontzen2012,DiCintio2014a,Tollet2016,Benitez-Llambay2019}.
A third scenario (SIDM) argues that dark matter self-interactions may
reduce the central DM densities relative to CDM, creating cores. As in
BICC, cusps may be re-formed in galaxies where baryons are gravitationally
important enough to deepen substantially the central potential
\citep[see, e.g.,][for a recent review]{Tulin2018}.

We have analyzed cosmological simulations of these two scenarios and
find that, although they both show promise explaining systems with
cores, neither reproduces the observed diversity in full
detail. Indeed, both scenarios have difficulty reproducing an
intriguing feature of the observed diversity, namely the existence of
galaxies with fast-rising rotation curves where the gravitational
effects of baryons in the inner regions is unimportant. They also face
difficulty explaining slowly-rising rotation curves where baryons
actually dominate in the inner regions, which are also present in the
observational sample we analyze.

We argue that these issues present a difficult problem for {\it any}
scenario where most halos are expected to develop a sizable core and
where baryons are supposed to be responsible for the observed
diversity. This is especially so because the relation between baryon
surface density and rotation curve shape is quite weak in the dwarf
galaxy regime, and thus unlikely to drive the diversity. We emphasize
that, strictly speaking, this conclusion applies only to the
particular implementations of BICC and SIDM we have tested here. These
are by no means the only possible realizations of these scenarios, and
it is definitely possible that further refinements may lead to
improvements in their accounting of the rotation curve diversity.

Our conclusions regarding
a second scenario, SIDM, may seem at odds with recent work that
reports good agreement between SIDM predictions and dwarf galaxy
rotation curves \citep[see; e.g., the recent preprint of][which
appeared as we were readying this paper for submission, and references
therein]{Kaplinghat2019}. That work, however, was meant to address
whether observed rotation curves {\it can} be reproduced by adjusting
the SIDM halo parameters freely in the fitting procedure, with
promising results. Our analysis, on the other hand, explores whether
the observed galaxies, if placed in average (random) SIDM halos, would
exhibit the observed diversity. Our results do show, in agreement with
earlier work, that SIDM leads to a wide distribution of rotation curve
shapes. However they also highlight the fact that outliers, be they
large cores or cuspy systems, are not readily accounted for in this
scenario, an issue that was also raised by
\citet{Creasey2017}. Whether this is a critical flaw of the SIDM
scenario, or just signals the need for further elaboration, is still
unclear.

In 
a third scenario the diversity is caused by variations in the baryonic
contribution to the acceleration in the inner regions, which is linked
to rotation velocities through the ``mass discrepancy-acceleration
relation'' \citep[MDAR;][]{McGaugh2016}. Earlier work has already
shown that such variations may indeed induce substantial diversity in
rotation curve shapes \citep{Ghari2019}. However, the MDAR is
incompatible with the trend between shape and baryonic importance and,
in particular, with the existence of galaxies with slowly-rising
rotation curves where baryons are important in the inner
regions. Reconciling these with the MDAR requires assuming that the
inner rotation velocities in those galaxies are in error. This is an
unappealing solution, since the MDAR was actually derived from such
data in the first place.

We end by noting that the rather peculiar relation between inner
baryon dominance and rotation curve shapes could be naturally
explained if non-circular motions were a driving cause of the
diversity. For this scenario to succeed, however, it would need to
explain why such motions affect solely low surface brightness
galaxies, the systems where the evidence for ``cores'' is most
compelling. Further progress in this regard would require a detailed
reanalysis of the data to uncover evidence for non-circular motions,
and a clear elaboration of the reason why non-circular motions do not
affect massive, high surface brightness galaxies.  Until then, we
would argue that the dwarf galaxy rotation curve diversity problem
remains, for the time being, open.

\section*{Acknowledgements}
We thank the anonymous referee whose suggestions have helped to improve this paper.
We also thank the NIHAO collaboration for kindly sharing their simulation
data, and
 Manoj Kaplinghat for useful comments.
 ISS and AR are supported by the Arthur B. McDonald Canadian
Astroparticle Physics Research Institute. JFN is a Fellow of the
Canadian Institute for Advanced Research and acknowledges useful
discussions with Laura Sales.  
MRL and AF had support from Marie Curie COFUND-Durham Junior 
Research Fellowship under EU grant agreement no. 609412.
MRL is funded by a Grant of Excellence from the Icelandic Research Fund (grant
number 173929-051).  
AF and CSF are supported by the Science and Technology Facilities Council (STFC)
 [grant numbers ST/F001166/1, ST/I00162X/1,ST/P000541/1].
KO received support from VICI grant 016.130.338
of the Netherlands Foundation for Scientific Research (NWO).  
AR is supported by the European Research Council (ERC-StG-716532-PUNCA) and
the STFC (ST/N001494/1). This work used the DiRAC@Durham facility
managed by the Institute for Computational Cosmology on behalf of the
STFC DiRAC HPC Facility (www.dirac.ac.uk). The equipment was funded by
BEIS capital funding via STFC capital grants ST/K00042X/1,
ST/P002293/1, ST/R002371/1 and ST/S002502/1, Durham University and
STFC operations grant ST/R000832/1. DiRAC is part of the National
e-Infrastructure.  ISS dedicates this work to her Abuelita, who was
fascinated by the Universe and outer space.


\bibliographystyle{mn2e}
\bibliography{archive}

\appendix

\section{SIDM transformation of LCDM halos}
\label{SecSIDMCDMTransf}

\subsection{Analytical model}
\label{SecSIDMCDMAnMod}

LCDM halos are well-characterized by a Navarro-Frenk-White (NFW) density
profile \citep{Navarro1996a,Navarro1997},
\begin{equation}
  \rho(r)={\rho_s \over (r/r_s)(1+r/r_s)^2},
  \label{EqNFW}
\end{equation}
which is fully specified by two parameters; e.g., a scale
density, $\rho_s$, and a scale radius, $r_s$ or, alternatively, by a
maximum circular velocity, $V_{\rm max}$ and the radius at which it is
achieved, $r_{\rm max}$. The two radial scales are related by
$r_{\rm max}=2.16\, r_s$.  The NFW profile has a $\rho \propto r^{-1}$ central cusp
where the velocity dispersion of dark matter particles (assuming
isotropy) decreases
steadily towards the centre, from a maximum value, $\sigma_{\rm
  max}=0.66\, V_{\rm max}$, that occurs at a radius $r_{\rm
  \sigma,max}=0.76\, r_s$. The radial dependence of the NFW velocity
dispersion has no simple algebraic form, but may be computed
numerically by assuming equilibrium and solving Jeans' equations
\citep[see; e.g., eq.~24 in][]{More2009}.

\begin{figure}
\includegraphics[width=0.7\linewidth]{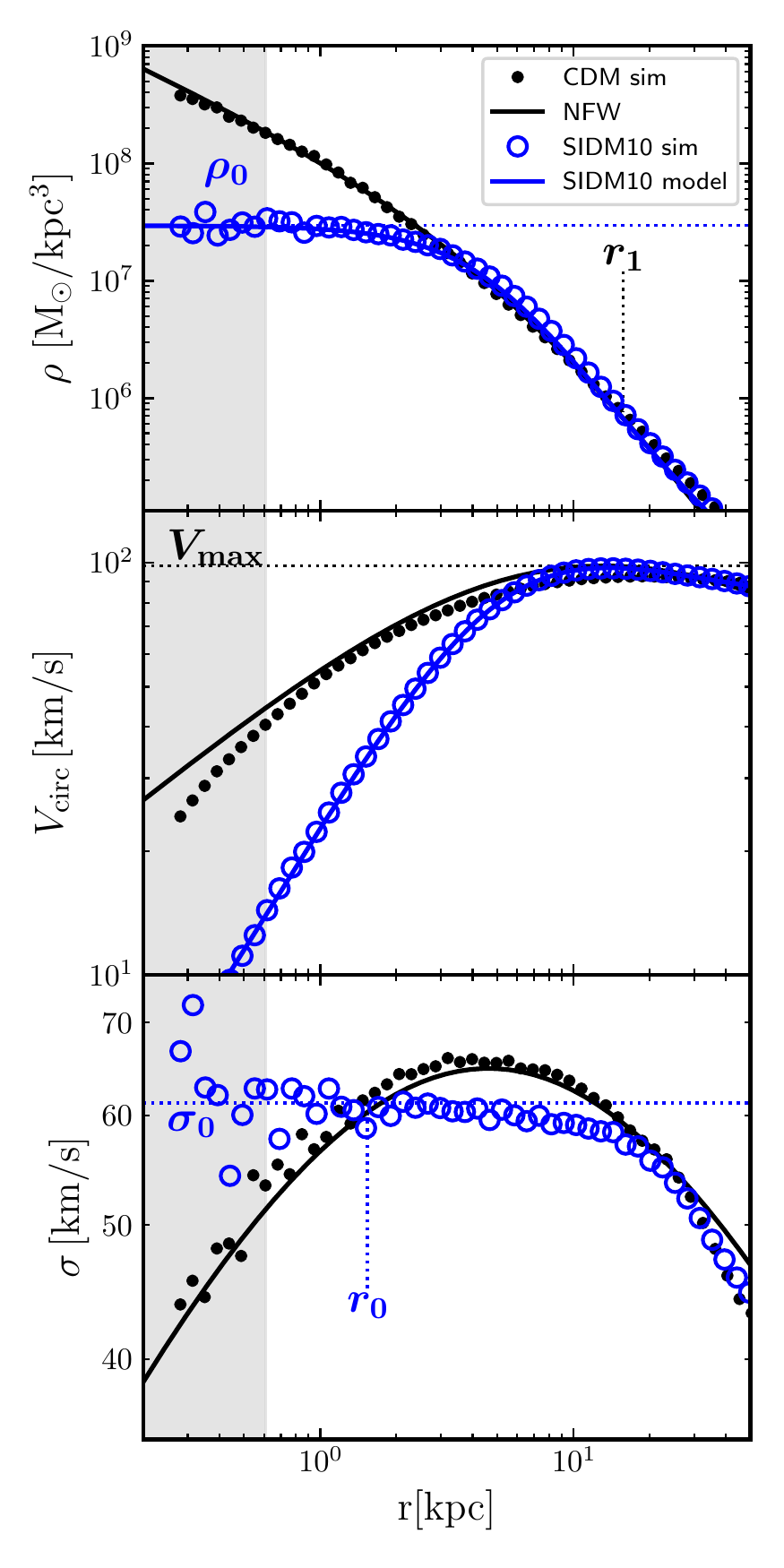}
\centering
\caption{Density (top), circular velocity (middle) and radial velocity
  dispersion (bottom) profiles for a low mass LCDM halo (black) and
  its SIDM counterpart evolved assuming $s_{\rm si}=10$ cm$^2$/g
  (blue). Thin solid black lines correspond to the best NFW fit to the
  LCDM halo profiles. Thin solid blue lines indicate the best fitting
  SIDM profile, computed as described in Sec.~\ref{SecSIDMCDMAnMod}. Some of
  the characteristic halo parameters are indicated in the profiles,
  such the central SIDM density, $\rho_0$, core radius, $r_{0}$,
  and central velocity dispersion, $\sigma_0$. The characteristic
  $r_1$ radius (eq.~\ref{EqR1}) and maximum velocity, $V_{\rm max}$,
  of the NFW fit are also indicated. The grey area indicates the
  region inside the \citet{Power2003} convergence radius.}
\label{FigCDMTransf}
\end{figure}

The solid black line in the top panel of Fig.~\ref{FigCDMTransf}
shows the NFW fit to the density profile of an LCDM halo of virial
mass $M_{200}=1.2 \times 10^{11}\, M_\odot$ (solid black circles)
selected from the AP-L1 dark-matter-only run. The fit has
concentration $c=r_{200}/r_s=17.3$; $r_s=6.06$ kpc; and
$\rho_s=5.62\times 10^6\, M_\odot/$kpc$^3$.  Assuming isotropy, the
NFW fit may be used to predict the velocity dispersion profile, which
is shown in the bottom panel of Fig.~\ref{FigCDMTransf}.

The main effect of self-interactions is to ``transfer heat'' from the
outside in,  ``thermalizing'' in the process the inner velocity
dispersion profile. This is demonstrated by the blue circles in
the bottom-panel of Fig.~\ref{FigCDMTransf}, which corresponds to the
same LCDM halo, but evolved with the dark-matter-only SIDM code
(Sec.~\ref{SecSIDMSim}), with $s_{\rm si}=10$ cm$^2$/g.

The isothermal region of the SIDM halo extends out to roughly $\sim
10$ kpc, which is comparable to $r_1$, the characteristic radius
where, for the assumed cross section, the local density in the LCDM
halo is such that one interaction is expected per Hubble time;
\begin{equation}
\left< s_{\rm si}  v_{\rm rel}\right> \rho_{\rm CDM}(r_1)\, t_0=1,
\label{EqR1}
\end{equation}
where, for the halo shown in Fig.~\ref{FigCDMTransf} we have assumed
$v_{\rm rel}=V_{\rm max}/\sqrt{2}$.

The mass profile of the SIDM halo may be derived (assuming isotropy
and spherical symmetry) by solving the hydrostatic equilibrium equation;
\begin{equation}
{d(\rho \sigma^2) \over dr}=-\rho {d\Phi \over dr}=-\rho {GM(r) \over r^2}=-\rho {V_c^2 \over r},
\label{EqHydro1}
\end{equation}
which, for $\sigma=\sigma_0=$ constant, may be written as
\begin{equation}
{d \ln \rho/\rho_0 \over dr}=-{1 \over r} {V_c^2 \over \sigma_0^2},
\label{EqHydro2}\end{equation}
where $\rho_0$ is a reference density.
We can multiply
eq.~\ref{EqHydro2} by $r^2$ and differentiate to get
\begin{equation}
{d \over dr} \left( r^2 {d\ln \rho/\rho_0 \over dr}  \right) = -
{d\over dr} \left( r {V_c^2 \over \sigma_0^2}\right)= -4\pi G
{\rho r^2 \over \sigma_0^2},
\label{EqHydro3}
\end{equation}
were we have used $d(rV_c^2)/dr=4\pi G \rho r^2$.

Differentiating Eq.~\ref{EqHydro3}, and defining $y\equiv
\ln(\rho/\rho_0)$, we have
\begin{equation}
r^2 {d^2y \over dr} + 2r {dy \over dr} +{4\pi G\rho_0 \over
  \sigma_0^2} e^y r^2=0,
\label{EqHydro4}
\end{equation}
where $r_{0}^2=4\pi G\rho_0 /  \sigma_0^2$ defines a
characteristic ``core'' radius, $r_{0}$. Expressing radii in units
of the core, $x=r/r_{0}$, we may rewrite Eq.~\ref{EqHydro4} as
\begin{equation}
  {d^2y \over dx^2}=-{2\over x}{dy \over dx}-e^y
\label{EqHydro5}
\end{equation}

Note that, in principle, this equation permits a family of solutions
for $\rho(r)$, many of which have cuspy inner profiles. (A simple
example is, of course, the singular isothermal sphere.)  Integrating
eq.~\ref{EqHydro5} numerically, therefore, requires that appropriate
boundary conditions are set.

Motivated by the density profile of the numerical SIDM halo shown in
the top panel of Fig.~\ref{FigCDMTransf}, we may set the conditions
$y(0)=0$ (i.e., finite central density) and $dy/dx(r=0)=0$ (i.e., a
flat density slope at the centre or ``constant density core'') to solve
for $y(x)$.

This solution gives a unique $\rho(r)$ profile that may be scaled for
any particular pair of values chosen from $\rho_0$, $r_{0}$, and
$\sigma_0$. The thin blue lines in the top two panels of
Fig.~\ref{FigCDMTransf} show one of these profiles, and demonstrates
that it provides an excellent fit to the actual inner density
profile of simulated SIDM halos. This analytical model is similar to that 
of \citet{Kaplinghat2014}.

\subsection{Empirical relations}
\label{SecSIDMCDMEmpir}

\begin{figure}
\includegraphics[width=\linewidth]{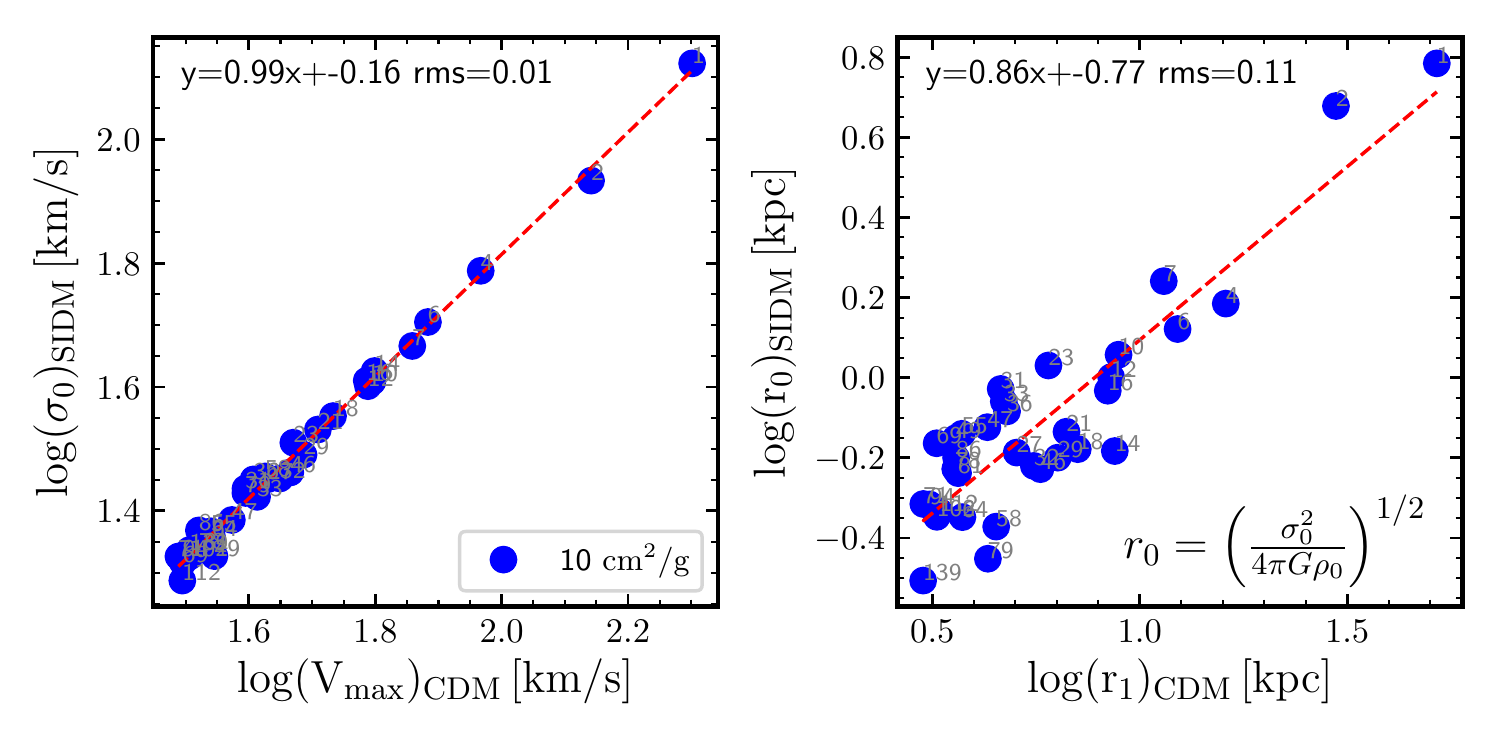}
\centering
\caption{ Empirical relations between the characteristic parameters of
  CDM and SIDM ($s_{\rm si}=10$ cm$^2$/g) matched halo pairs in the
  APOSTLE L1-V1 simulation. \textit{Left:} Central SIDM velocity
  dispersion, $\sigma_0$, versus LCDM $V_{\rm max}$. \textit{Right:}
  Characteristic SIDM core radius  $r_{0}$ versus the characteristic LCDM radius,
  $r_1$, given by Eq.~\ref{EqR1}.}
\label{FigCDMSIDMScatter}
\end{figure}

As decribed above, describing the inner mass profile of an SIDM halo
requires that two parameters be specified. These parameters are
expected to be closely related to the NFW parameters of its
corresponding LCDM halo. We show this in Fig.~\ref{FigCDMSIDMScatter}
for matching pairs of halos identified in dark-matter-only LCDM and
SIDM runs of one of the AP-L1 volumes. The SIDM run assumes
$s_{\rm si}=10$ cm$^2$/g.

Fig.~\ref{FigCDMSIDMScatter} shows the relation between the maximum
circular velocity of the LCDM halo and the central velocity dispersion
of its SIDM counterpart ($\sigma_0$, left panel), as well as the
relation between the LCDM characteristic radius $r_1$ and the
corresponding SIDM core radius, $r_{0}$. These two relations, together
with their scatter, may be used to generate a population of SIDM halos
using the LCDM $M_{200}(c)$ relation (and its scatter), since the mass
and concentration of an NFW halo fully specify $r_1$ and
$V_{\rm max}$. For Planck-normalized LCDM, and $s_{\rm si}=10$ cm$^2$/g,
\begin{equation}
 r_1= \biggl({V_{\rm max} \over 14.5 \, {\rm km/s}}\biggr)^{1.5} {\rm kpc},
\end{equation}
provides a good approximation to this relation.

The blue bands shown in Fig.~\ref{FigVfidVmaxSIDM} were computed this
way, after transforming thousands of Planck-normalized LCDM halos with
$M_{200}$ and $c$ sampled from the mass-concentration relation of
\citet{Ludlow2016}. We emphasize that the relations shown in
Fig.~\ref{FigCDMSIDMScatter} are sensitive to the assumed value of
$s_{\rm si}=10$ cm$^2$/g.

\subsection{Including baryons in SIDM halos}
\label{SecSIDMCDMIncBar}

The formulation described in Sec.~\ref{SecSIDMCDMAnMod} may be easily modified
to account for an additional (baryonic) mass component. In this case,
the total gravitational potential is modified so that the circular
velocity term in Eq.~\ref{EqHydro1} is split between the dark
matter and baryon contributions as $V_c^2=V_{\rm dm}^2+V_{\rm
  bar}^2$. Carrying this change through the derivation yields a
modified version of Eq.~\ref{EqHydro5};
\begin{equation}
{d^2y \over dx^2}=-{2\over x}{dy \over dx}-e^y-{1\over x^2}{d \over
    dx} \left( x{V_{\rm bar}^2 \over \sigma_0^2}\right).
\label{EqHydro6}  
\end{equation}

This allows the mass profile of the baryonic component, specified by
$V_{\rm bar}(r)$, to be readily included when computing the dark
matter profile of the SIDM halo. In order to set the appropriate
boundary conditions, we note that isothermal distributions generally
satisfy, from inspection of eq.~\ref{EqHydro1}, that
\begin{equation}
  \rho \propto e^{-\Phi/\sigma_0^2}.
  \label{EqIso}
\end{equation}
This implies that the deeper the central potential becomes because of
the baryonic component, the higher the central dark matter density of
the SIDM halo is expected to be.

In practice, our modeling proceeds as follows. Given an SIDM halo
characterized by a central density and core radius, which fully
specify its circular velocity profile, $V_{\rm dm}^2(r)$, we would
like to compute how it would be modified by the addition of a baryonic
component, specified by $V_{\rm bar}^2(r)$ over the radial range
($0$, $r_{\rm last}$), where the rotation curve has been measured.  The
gravitational potential change between these two radii is given by
\begin{equation}
\delta \Phi_{0l}={\Phi(r_{\rm last})-\Phi(0) \over
  \sigma_0^2}=\int_0^{r_{\rm last}} {V_{\rm dm}^2+V_{\rm bar}^2 \over
  \sigma_0^2} {dr' \over r'},
\end{equation}
which, in turn, should set the drop in dark matter density from the
centre to $r_{\rm last}$; i.e.,
$\rho_{\rm dm}(0)/\rho_{\rm dm}(r_{\rm last})=\exp({\delta
  \Phi_{0l}})$, according to Eq.~\ref{EqIso}.

We impose this condition when solving Eq.~\ref{EqHydro6} to find the
dark matter density profile in the presence of baryons. The final dark
matter profile is renormalized by the universal baryon fraction to
take into account the fact that the SIDM halo parameters computed as
described in Sec.~\ref{SecSIDMCDMEmpir} were computed from a
dark-matter-only simulation. We also make a small adjustment to
account for the fact that the potential wells of low mass halos in
cosmological hydrodyamical simulations are systematically less deep
than in dark-matter-only runs \citep[see; e.g.,][]{Sawala2016}. These
adjustments have a minor effect on our results, but help to
reconcile our analytical model with the results of the SIDM cosmological
hydrodynamical run shown in the bottom-right panels of
Fig.~\ref{FigVfidVmaxSIDM} and Fig.~\ref{FigEtaRotEtaBarSIDM}.

\begin{figure}
\includegraphics[width=\linewidth]{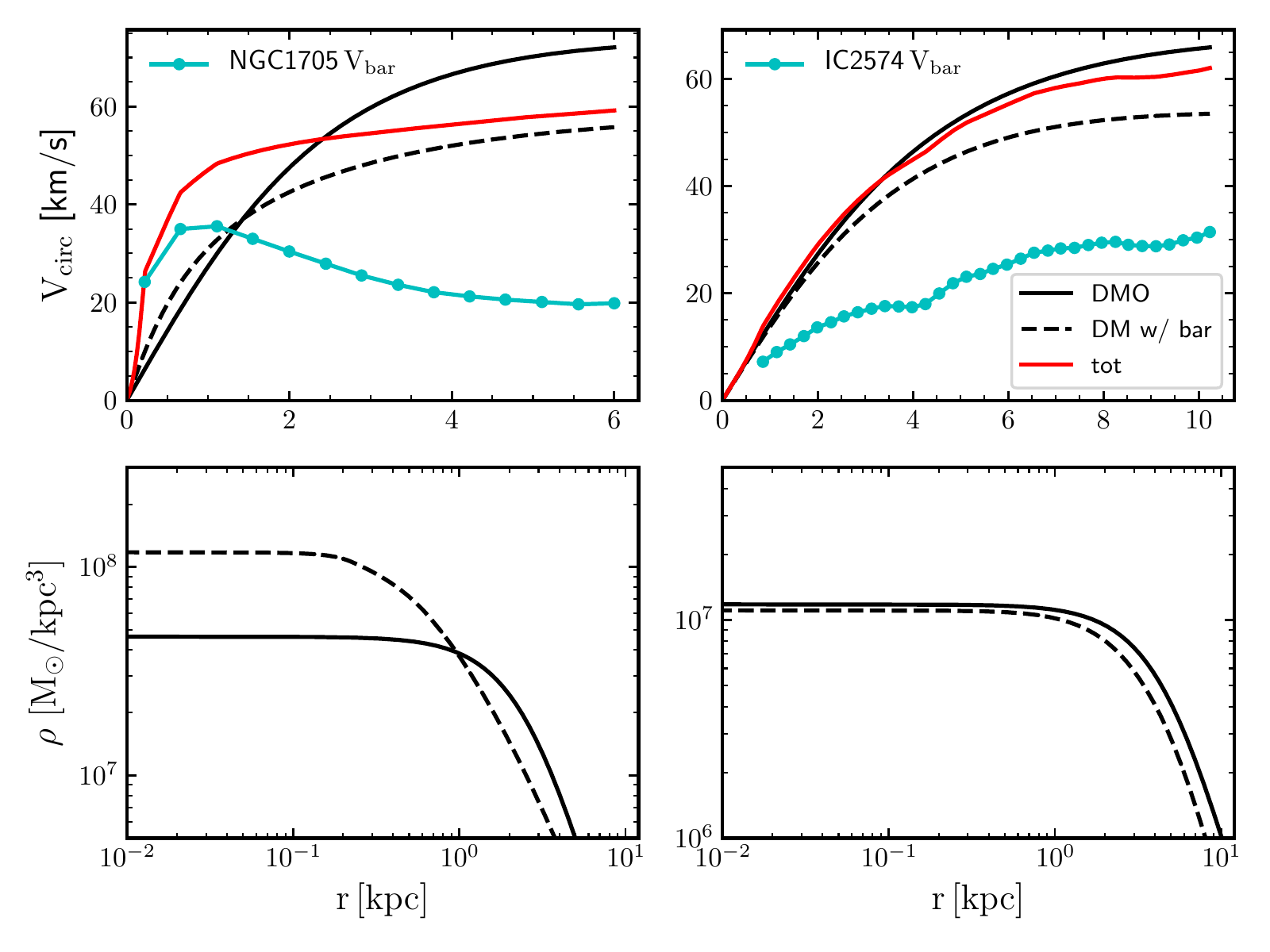}
\centering
\caption{Baryon-induced transformation of an SIDM halo, according to
  the model described in Sec.~\ref{SecSIDMCDMIncBar}. Two examples are given,
  where the baryonic mass profiles of NGC 1075 (left) and IC 2574
  (right) are added to randomly selected SIDM halos with matching
  $V_{\rm max}$ ($73$ km/s for NGC 1075, and $68$ km/s for IC
  2574). The dark-matter-only SIDM halo profiles are shown with solid
  lines. The resulting profiles induced by the baryonic component and
  the rescaling/normalization procedure described in
  Sec.~\ref{SecSIDMCDMIncBar} are shown with dashed lines. Note that the dark
  matter profile is less affected in the case of IC 2574 than for NGC
  1075 because the baryonic mass distribution is more extended in the
  former and has a smaller effect on the central gravitational
  potential.}
\label{FigCDMSIDMModel}
\end{figure}

For illustration, we show two examples of our procedure in
Fig.~\ref{FigCDMSIDMModel}. This illustrates the effect on the inner
dark matter profile of two rather different baryonic distributions;
that of NGC 1075 (left) and that of IC 2574 (right), when added to two
randomly drawn SIDM halos with maximum velocities matching the
observed values of each of these galaxies. The solid black lines
indicate the dark-matter-only SIDM profile; the dashed black lines are
the resulting profiles after the baryons have been added. Note that
the dashed-line profiles have less dark matter in total because of the
baryonic mass renormalization described in the preceding
paragraph. The addition of baryons leads, as expected, to dark matter
profiles that are denser near the centre and that drop more sharply
than their dark-matter-only counterparts.


\begin{table*}
\centering
\scriptsize
\caption{Observational data used in this work.
Column 2 provides the sample  the galaxy belongs to: 'S'  \citep[SPARC,][]{Lelli2016}, 'LT' \citep[LITTLE THINGS][]{Oh2015}, 'TH' \citep[THINGS][]{deBlok2008}, 'A' \citep{Adams2014}, 'R' \citep{Relatores2019}. 
}

\begin{tabular}{ l l l l l l l l l }

\toprule
Name & Sample & $V_{\rm max}$ [km/s] & $V_{\rm b,max}$ [km/s]  & $V_{\rm fid}$  [km/s] & $V_{\rm b,fid}$ [km/s]  & $M_{\rm bar}$ [$M_\odot$] &  $r_{\rm b,half}$ [kpc] & $M_{200}$ [$M_\odot$]  \\

\midrule
\midrule
CamB    &    S    &    20.10    &    13.85    &    6.85    &    10.08    &    5.35e+07    &    0.86    &    1.23e+09 \\
D512-2    &    S    &    37.20    &    16.34    &    24.04    &    11.80    &    2.70e+08    &    1.94    &    8.50e+09 \\
D564-8    &    S    &    25.00    &    9.24    &    11.17    &    6.66    &    5.51e+07    &    1.50    &    2.43e+09 \\
D631-7    &    S    &    58.50    &    24.51    &    29.50    &    23.54    &    4.84e+08    &    2.03    &    3.55e+10 \\
DDO064    &    S    &    46.90    &    25.29    &    36.67    &    16.28    &    3.59e+08    &    1.86    &    1.76e+10 \\
DDO154    &    S    &    48.20    &    18.73    &    27.37    &    13.90    &    3.92e+08    &    2.71    &    1.92e+10 \\
DDO161    &    S    &    67.50    &    35.35    &    32.14    &    25.01    &    2.11e+09    &    5.59    &    5.58e+10 \\
DDO168    &    S    &    55.00    &    25.90    &    35.00    &    23.08    &    6.45e+08    &    2.12    &    2.92e+10 \\
ESO079-G014    &    S    &    178.00    &    100.77    &    109.73    &    64.21    &    3.00e+10    &    9.46    &    1.21e+12 \\
ESO116-G012    &    S    &    112.00    &    52.33    &    83.47    &    47.73    &    3.59e+09    &    4.07    &    2.78e+11 \\
ESO444-G084    &    S    &    63.10    &    21.89    &    52.73    &    21.31    &    2.15e+08    &    1.38    &    4.51e+10 \\
ESO563-G021    &    S    &    321.00    &    200.92    &    286.19    &    176.36    &    1.88e+11    &    12.00    &    7.95e+12 \\
F565-V2    &    S    &    83.10    &    30.16    &    38.28    &    15.53    &    1.21e+09    &    6.13    &    1.08e+11 \\
F568-3    &    S    &    120.00    &    52.15    &    66.15    &    29.09    &    8.42e+09    &    10.60    &    3.46e+11 \\
F568-V1    &    S    &    118.00    &    44.75    &    91.59    &    28.75    &    5.23e+09    &    9.82    &    3.28e+11 \\
F571-8    &    S    &    144.00    &    81.10    &    93.14    &    70.96    &    7.45e+09    &    2.04    &    6.18e+11 \\
F574-1    &    S    &    99.70    &    46.48    &    68.23    &    29.70    &    7.96e+09    &    8.73    &    1.92e+11 \\
F583-1    &    S    &    86.90    &    41.35    &    44.55    &    13.55    &    3.32e+09    &    10.70    &    1.24e+11 \\
F583-4    &    S    &    69.90    &    31.15    &    43.05    &    23.48    &    1.71e+09    &    4.12    &    6.23e+10 \\
IC2574    &    S    &    67.50    &    32.02    &    19.54    &    12.57    &    1.89e+09    &    7.24    &    5.58e+10 \\
IC4202    &    S    &    250.00    &    181.45    &    231.32    &    165.39    &    1.06e+11    &    9.07    &    3.58e+12 \\
KK98-251    &    S    &    34.60    &    20.25    &    16.35    &    10.59    &    1.95e+08    &    1.83    &    6.76e+09 \\
NGC0024    &    S    &    110.00    &    52.54    &    106.73    &    51.90    &    2.84e+09    &    2.53    &    2.62e+11 \\
NGC0055    &    S    &    87.40    &    50.19    &    43.53    &    33.96    &    4.40e+09    &    6.74    &    1.26e+11 \\
NGC0100    &    S    &    91.20    &    40.75    &    56.83    &    36.27    &    4.26e+09    &    3.72    &    1.45e+11 \\
NGC0247    &    S    &    108.00    &    53.09    &    66.63    &    29.91    &    5.99e+09    &    8.02    &    2.48e+11 \\
NGC0289    &    S    &    194.00    &    160.25    &    174.05    &    156.00    &    7.26e+10    &    16.54    &    1.60e+12 \\
NGC0300    &    S    &    97.00    &    41.87    &    66.19    &    35.13    &    2.71e+09    &    4.56    &    1.76e+11 \\
NGC0801    &    S    &    238.00    &    199.22    &    233.82    &    195.75    &    1.87e+11    &    15.86    &    3.06e+12 \\
NGC0891    &    S    &    234.00    &    265.08    &    224.00    &    214.65    &    7.51e+10    &    2.79    &    2.90e+12 \\
NGC1003    &    S    &    115.00    &    59.35    &    74.39    &    55.38    &    1.12e+10    &    12.36    &    3.02e+11 \\
NGC1090    &    S    &    176.00    &    121.39    &    154.59    &    108.35    &    4.77e+10    &    8.68    &    1.17e+12 \\
NGC1705    &    S    &    73.20    &    35.42    &    72.84    &    29.77    &    4.51e+08    &    0.95    &    7.21e+10 \\
NGC2366    &    S    &    53.70    &    28.57    &    32.98    &    20.62    &    9.79e+08    &    3.25    &    2.71e+10 \\
NGC2403    &    S    &    136.00    &    75.85    &    117.48    &    75.48    &    9.28e+09    &    5.07    &    5.15e+11 \\
NGC2683    &    S    &    212.00    &    171.90    &    211.40    &    169.48    &    4.21e+10    &    3.45    &    2.12e+12 \\
NGC2841    &    S    &    323.00    &    231.30    &    322.93    &    214.81    &    1.07e+11    &    7.98    &    8.11e+12 \\
NGC2903    &    S    &    216.00    &    213.11    &    207.86    &    176.97    &    4.43e+10    &    2.58    &    2.25e+12 \\
NGC2915    &    S    &    86.50    &    35.40    &    77.83    &    27.64    &    9.96e+08    &    4.97    &    1.22e+11 \\
NGC2955    &    S    &    276.00    &    252.60    &    245.59    &    246.68    &    1.98e+11    &    8.12    &    4.91e+12 \\
NGC2998    &    S    &    214.00    &    157.58    &    204.14    &    150.74    &    1.07e+11    &    11.13    &    2.18e+12 \\
NGC3109    &    S    &    67.30    &    21.72    &    36.06    &    15.30    &    7.31e+08    &    3.92    &    5.52e+10 \\
NGC3198    &    S    &    157.00    &    101.32    &    118.91    &    85.11    &    3.36e+10    &    10.54    &    8.13e+11 \\
NGC3521    &    S    &    220.00    &    200.33    &    214.79    &    174.16    &    4.79e+10    &    2.99    &    2.38e+12 \\
NGC3726    &    S    &    169.00    &    131.38    &    119.56    &    111.65    &    4.37e+10    &    6.73    &    1.03e+12 \\
NGC3741    &    S    &    51.60    &    14.28    &    27.05    &    11.19    &    2.56e+08    &    3.79    &    2.39e+10 \\
NGC3769    &    S    &    126.00    &    98.81    &    106.57    &    98.43    &    1.67e+10    &    4.28    &    4.04e+11 \\
NGC3877    &    S    &    171.00    &    145.67    &    157.15    &    109.76    &    3.82e+10    &    6.22    &    1.07e+12 \\
NGC3893    &    S    &    194.00    &    152.06    &    193.30    &    149.89    &    3.70e+10    &    3.83    &    1.60e+12 \\
NGC3917    &    S    &    138.00    &    80.96    &    104.05    &    57.20    &    1.35e+10    &    6.35    &    5.40e+11 \\
NGC3953    &    S    &    224.00    &    176.03    &    211.67    &    159.61    &    7.44e+10    &    7.33    &    2.52e+12 \\
NGC3972    &    S    &    134.00    &    79.21    &    105.02    &    62.09    &    8.79e+09    &    4.75    &    4.91e+11 \\
NGC4010    &    S    &    129.00    &    77.65    &    86.12    &    68.91    &    1.24e+10    &    5.15    &    4.35e+11 \\
NGC4013    &    S    &    198.00    &    166.75    &    197.26    &    161.12    &    4.35e+10    &    5.49    &    1.70e+12 \\
NGC4051    &    S    &    161.00    &    141.36    &    131.17    &    103.88    &    5.12e+10    &    6.35    &    8.81e+11 \\
NGC4088    &    S    &    182.00    &    189.42    &    161.40    &    170.20    &    6.46e+10    &    5.21    &    1.30e+12 \\
NGC4100    &    S    &    195.00    &    147.99    &    185.05    &    132.10    &    3.38e+10    &    5.13    &    1.62e+12 \\
NGC4138    &    S    &    195.00    &    171.38    &    193.22    &    142.04    &    2.40e+10    &    2.61    &    1.62e+12 \\
NGC4157    &    S    &    201.00    &    182.02    &    194.60    &    174.37    &    6.38e+10    &    4.96    &    1.79e+12 \\
NGC4183    &    S    &    115.00    &    66.98    &    83.39    &    43.71    &    1.01e+10    &    7.10    &    3.02e+11 \\
NGC4217    &    S    &    191.00    &    234.50    &    177.25    &    205.45    &    4.61e+10    &    2.69    &    1.52e+12 \\
NGC4559    &    S    &    124.00    &    83.97    &    98.55    &    78.38    &    1.74e+10    &    7.55    &    3.84e+11 \\
NGC5033    &    S    &    225.00    &    241.63    &    224.72    &    185.97    &    7.03e+10    &    3.56    &    2.56e+12 \\
NGC5055    &    S    &    206.00    &    198.71    &    201.57    &    196.38    &    9.21e+10    &    5.53    &    1.93e+12 \\
NGC5371    &    S    &    242.00    &    213.77    &    214.79    &    173.23    &    1.85e+11    &    11.10    &    3.23e+12 \\
NGC5585    &    S    &    92.30    &    44.82    &    62.65    &    40.75    &    3.71e+09    &    4.59    &    1.50e+11 \\
NGC5907    &    S    &    235.00    &    165.49    &    218.01    &    139.96    &    1.16e+11    &    11.24    &    2.94e+12 \\
NGC5985    &    S    &    305.00    &    165.00    &    288.80    &    140.14    &    1.20e+11    &    12.81    &    6.75e+12 \\
NGC6015    &    S    &    166.00    &    106.26    &    151.16    &    106.20    &    2.38e+10    &    5.11    &    9.71e+11 \\
NGC6195    &    S    &    258.00    &    235.77    &    233.45    &    235.70    &    2.23e+11    &    10.62    &    3.96e+12 \\
NGC6503    &    S    &    121.00    &    101.33    &    119.67    &    91.08    &    8.74e+09    &    2.16    &    3.55e+11 \\
NGC6674    &    S    &    291.00    &    177.58    &    285.98    &    175.67    &    1.50e+11    &    10.19    &    5.81e+12 \\
NGC6946    &    S    &    181.00    &    195.85    &    176.12    &    149.58    &    4.06e+10    &    4.39    &    1.28e+12 \\
NGC7331    &    S    &    257.00    &    275.82    &    257.00    &    255.45    &    1.40e+11    &    4.40    &    3.91e+12 \\
NGC7793    &    S    &    116.00    &    72.47    &    102.82    &    72.45    &    4.67e+09    &    2.55    &    3.11e+11 \\
NGC7814    &    S    &    265.00    &    290.19    &    226.72    &    163.25    &    3.87e+10    &    1.66    &    4.31e+12 \\
PGC51017    &    S    &    20.50    &    18.81    &    20.27    &    12.47    &    3.45e+08    &    1.58    &    1.31e+09 \\
UGC00128    &    S    &    134.00    &    55.39    &    78.98    &    35.06    &    1.59e+10    &    18.83    &    4.91e+11 \\
UGC00191    &    S    &    83.85    &    42.19    &    71.49    &    30.11    &    2.79e+09    &    5.41    &    1.11e+11 \\
UGC00731    &    S    &    74.00    &    37.19    &    46.52    &    10.43    &    2.56e+09    &    7.30    &    7.46e+10 \\
\end{tabular}
\label{TabObsData}
\end{table*}

\begin{table*}
\centering
\scriptsize
\contcaption{Table \ref{TabObsData}. }

\begin{tabular}{ l l l l l l l l l }

\toprule
Name & Sample & $V_{\rm max}$ [km/s] & $V_{\rm b,max}$ [km/s]  & $V_{\rm fid}$  [km/s] & $V_{\rm b,fid}$ [km/s]  & $M_{\rm bar}$ [$M_\odot$] &  $r_{\rm b,half}$ [kpc] & $M_{200}$ [$M_\odot$]  \\

\midrule
\midrule
UGC00891    &    S    &    63.75    &    26.57    &    28.75    &    17.86    &    7.56e+08    &    3.50    &    4.65e+10 \\
UGC02259    &    S    &    90.00    &    36.92    &    76.15    &    29.49    &    1.52e+09    &    4.01    &    1.39e+11 \\
UGC02455    &    S    &    61.00    &    57.14    &    30.44    &    56.18    &    2.89e+09    &    1.93    &    4.05e+10 \\
UGC02487    &    S    &    383.00    &    251.56    &    376.62    &    250.17    &    2.69e+11    &    10.67    &    1.40e+13 \\
UGC02885    &    S    &    305.00    &    231.80    &    260.15    &    210.89    &    2.55e+11    &    21.76    &    6.75e+12 \\
UGC02916    &    S    &    218.00    &    237.95    &    209.71    &    197.65    &    9.30e+10    &    8.55    &    2.31e+12 \\
UGC02953    &    S    &    319.00    &    247.65    &    319.00    &    238.35    &    1.40e+11    &    5.31    &    7.79e+12 \\
UGC03205    &    S    &    237.00    &    174.00    &    236.00    &    173.49    &    6.97e+10    &    5.94    &    3.02e+12 \\
UGC03546    &    S    &    262.00    &    330.90    &    193.03    &    167.97    &    5.42e+10    &    2.52    &    4.16e+12 \\
UGC03580    &    S    &    131.00    &    97.26    &    96.63    &    84.30    &    1.24e+10    &    5.84    &    4.57e+11 \\
UGC04278    &    S    &    92.80    &    46.03    &    45.88    &    21.46    &    2.14e+09    &    5.70    &    1.53e+11 \\
UGC04305    &    S    &    37.30    &    33.87    &    24.47    &    19.17    &    1.29e+09    &    3.21    &    8.57e+09 \\
UGC04325    &    S    &    92.70    &    44.02    &    87.71    &    37.41    &    1.91e+09    &    3.14    &    1.52e+11 \\
UGC04499    &    S    &    74.30    &    39.87    &    49.06    &    27.20    &    2.24e+09    &    4.77    &    7.56e+10 \\
UGC05005    &    S    &    100.00    &    43.08    &    37.87    &    25.08    &    6.16e+09    &    15.24    &    1.94e+11 \\
UGC05253    &    S    &    248.00    &    238.24    &    246.00    &    208.09    &    1.08e+11    &    6.08    &    3.49e+12 \\
UGC05414    &    S    &    61.40    &    33.68    &    41.04    &    26.20    &    1.32e+09    &    2.52    &    4.13e+10 \\
UGC05716    &    S    &    74.70    &    31.57    &    49.52    &    20.96    &    1.75e+09    &    6.86    &    7.69e+10 \\
UGC05721    &    S    &    82.60    &    34.76    &    81.92    &    29.52    &    1.01e+09    &    3.33    &    1.06e+11 \\
UGC05750    &    S    &    78.90    &    41.46    &    24.89    &    15.88    &    3.13e+09    &    13.59    &    9.14e+10 \\
UGC05764    &    S    &    55.80    &    22.22    &    49.25    &    16.24    &    2.59e+08    &    2.08    &    3.05e+10 \\
UGC05829    &    S    &    68.60    &    37.43    &    34.69    &    12.54    &    1.64e+09    &    5.10    &    5.87e+10 \\
UGC05918    &    S    &    44.50    &    19.72    &    29.16    &    9.21    &    5.12e+08    &    3.04    &    1.49e+10 \\
UGC05986    &    S    &    116.00    &    56.04    &    98.38    &    51.27    &    5.89e+09    &    3.44    &    3.11e+11 \\
UGC06399    &    S    &    87.60    &    38.86    &    58.74    &    27.28    &    2.04e+09    &    4.54    &    1.27e+11 \\
UGC06446    &    S    &    84.90    &    38.55    &    64.44    &    24.16    &    2.33e+09    &    6.00    &    1.15e+11 \\
UGC06614    &    S    &    205.00    &    209.41    &    184.29    &    186.91    &    9.13e+10    &    13.15    &    1.90e+12 \\
UGC06667    &    S    &    85.70    &    30.25    &    54.81    &    12.29    &    1.77e+09    &    5.12    &    1.19e+11 \\
UGC06786    &    S    &    229.00    &    175.04    &    219.00    &    155.28    &    4.34e+10    &    4.47    &    2.71e+12 \\
UGC06787    &    S    &    276.00    &    271.81    &    243.26    &    170.76    &    5.58e+10    &    2.77    &    4.91e+12 \\
UGC06818    &    S    &    74.40    &    32.73    &    31.86    &    28.59    &    2.23e+09    &    3.73    &    7.59e+10 \\
UGC06917    &    S    &    111.00    &    56.30    &    79.59    &    43.25    &    6.11e+09    &    6.44    &    2.70e+11 \\
UGC06923    &    S    &    81.10    &    47.91    &    61.65    &    43.56    &    2.52e+09    &    2.88    &    9.97e+10 \\
UGC06930    &    S    &    109.00    &    60.56    &    78.67    &    41.51    &    8.77e+09    &    8.06    &    2.55e+11 \\
UGC06983    &    S    &    113.00    &    51.91    &    90.67    &    43.65    &    6.60e+09    &    7.81    &    2.86e+11 \\
UGC07089    &    S    &    79.10    &    42.06    &    40.58    &    29.31    &    3.41e+09    &    5.01    &    9.22e+10 \\
UGC07125    &    S    &    65.60    &    44.56    &    31.20    &    24.34    &    7.51e+09    &    10.46    &    5.10e+10 \\
UGC07151    &    S    &    76.20    &    44.10    &    64.66    &    36.63    &    1.96e+09    &    3.54    &    8.19e+10 \\
UGC07323    &    S    &    85.60    &    51.74    &    52.58    &    35.03    &    3.01e+09    &    3.91    &    1.18e+11 \\
UGC07399    &    S    &    106.00    &    33.87    &    91.53    &    32.51    &    1.57e+09    &    3.27    &    2.33e+11 \\
UGC07524    &    S    &    83.80    &    43.92    &    44.96    &    19.35    &    3.58e+09    &    7.73    &    1.11e+11 \\
UGC07559    &    S    &    32.10    &    19.76    &    18.91    &    13.18    &    2.79e+08    &    1.79    &    5.34e+09 \\
UGC07577    &    S    &    17.80    &    12.60    &    6.69    &    5.52    &    8.10e+07    &    1.10    &    8.39e+08 \\
UGC07603    &    S    &    64.00    &    27.55    &    58.22    &    27.15    &    5.31e+08    &    1.56    &    4.71e+10 \\
UGC07690    &    S    &    60.70    &    36.49    &    60.65    &    35.71    &    9.48e+08    &    1.59    &    3.99e+10 \\
UGC07866    &    S    &    33.10    &    18.42    &    23.18    &    12.75    &    2.19e+08    &    1.53    &    5.88e+09 \\
UGC08286    &    S    &    84.30    &    35.34    &    72.65    &    29.57    &    1.48e+09    &    3.94    &    1.13e+11 \\
UGC08490    &    S    &    80.10    &    35.71    &    76.29    &    35.55    &    1.47e+09    &    2.96    &    9.59e+10 \\
UGC08550    &    S    &    57.80    &    24.64    &    48.45    &    23.06    &    5.28e+08    &    3.96    &    3.41e+10 \\
UGC08699    &    S    &    202.00    &    214.73    &    170.49    &    125.86    &    3.01e+10    &    3.85    &    1.82e+12 \\
UGC08837    &    S    &    48.00    &    29.32    &    18.56    &    13.25    &    6.76e+08    &    2.82    &    1.90e+10 \\
UGC09037    &    S    &    160.00    &    116.33    &    102.52    &    107.36    &    5.97e+10    &    10.23    &    8.64e+11 \\
UGC09133    &    S    &    289.00    &    283.20    &    287.52    &    244.07    &    1.86e+11    &    8.06    &    5.69e+12 \\
UGC10310    &    S    &    73.20    &    40.86    &    52.31    &    24.84    &    2.46e+09    &    4.84    &    7.21e+10 \\
UGC11455    &    S    &    291.00    &    232.15    &    232.60    &    204.66    &    2.05e+11    &    10.36    &    5.81e+12 \\
UGC11820    &    S    &    84.45    &    38.77    &    49.23    &    20.19    &    3.11e+09    &    8.75    &    1.13e+11 \\
UGC12506    &    S    &    255.00    &    134.55    &    210.25    &    101.90    &    1.17e+11    &    18.78    &    3.82e+12 \\
UGC12632    &    S    &    73.20    &    36.66    &    44.04    &    18.21    &    2.97e+09    &    7.28    &    7.21e+10 \\
UGC12732    &    S    &    98.00    &    46.29    &    59.03    &    23.06    &    5.70e+09    &    11.78    &    1.82e+11 \\
UGCA442    &    S    &    57.80    &    24.49    &    34.31    &    18.27    &    4.20e+08    &    2.18    &    3.41e+10 \\
UGCA444    &    S    &    38.30    &    16.18    &    25.38    &    14.49    &    9.51e+07    &    1.19    &    9.32e+09 \\
U11707    &    A    &    103.74    &    32.33    &    63.19    &    22.91    &    2.24e+09    &    5.99    &    2.18e+11 \\
N2552    &    A    &    96.10    &    35.28    &    68.53    &    30.81    &    1.78e+09    &    3.43    &    1.71e+11 \\
wlm    &    LT    &    38.53    &    14.84    &    26.93    &    10.86    &    1.73e+08    &    2.22    &    9.49e+09 \\
ddo87    &    LT    &    56.63    &    18.79    &    22.83    &    4.92    &    5.85e+08    &    4.68    &    3.20e+10 \\
ddo50    &    LT    &    38.84    &    33.88    &    22.29    &    7.38    &    1.67e+09    &    4.62    &    9.74e+09 \\
ddo52    &    LT    &    61.72    &    22.52    &    38.71    &    10.70    &    6.30e+08    &    3.80    &    4.20e+10 \\
ngc1569    &    LT    &    39.28    &    38.68    &    23.57    &    20.45    &    1.12e+09    &    2.84    &    1.01e+10 \\
haro29    &    LT    &    43.49    &    12.00    &    29.82    &    9.69    &    1.68e+08    &    4.36    &    1.39e+10 \\
cvnidwa    &    LT    &    26.45    &    8.89    &    9.14    &    8.43    &    4.53e+07    &    1.68    &    2.91e+09 \\
ddo133    &    LT    &    46.73    &    18.84    &    32.36    &    11.48    &    2.87e+08    &    2.28    &    1.74e+10 \\
ic1613    &    LT    &    21.13    &    15.29    &    6.13    &    5.03    &    1.47e+08    &    1.82    &    1.44e+09 \\
ddo216    &    LT    &    18.91    &    8.45    &    12.69    &    6.90    &    1.73e+07    &    0.60    &    1.01e+09 \\
ddo126    &    LT    &    38.74    &    18.70    &    19.63    &    10.07    &    2.91e+08    &    2.44    &    9.66e+09 \\
NGC925    &    TH    &    114.50    &    75.56    &    54.41    &    46.03    &    1.73e+10    &    9.65    &    2.98e+11 \\
NGC3621    &    TH    &    159.20    &    118.03    &    130.41    &    116.23    &    3.48e+10    &    6.38    &    8.50e+11 \\
ngc4396    &    R    &    99.46    &    30.63    &    79.58    &    30.13    &    1.12e+09    &    2.71    &    1.91e+11 \\
ngc6106    &    R    &    124.75    &    68.67    &    114.06    &    68.32    &    5.46e+09    &    2.58    &    3.91e+11 \\
ugc4169    &    R    &    97.72    &    46.37    &    74.84    &    38.39    &    3.11e+09    &    3.65    &    1.80e+11 \\
ugc3371    &    R    &    64.01    &    15.95    &    36.79    &    10.54    &    3.98e+08    &    4.26    &    4.71e+10 \\
\bottomrule

\end{tabular}
\end{table*}

\label{lastpage}

\end{document}